\documentclass[journal]{IEEEtran}

\usepackage{amsmath,bbold, setspace, stfloats, dsfont,yfonts,amssymb,dsfont,epsfig,psfrag,amsthm,bm,multirow, graphicx,color,mathrsfs}

\usepackage{algpseudocode}
\usepackage{algorithm,eqparbox,array}
\usepackage{xspace}

\usepackage{caption2}

\newtheorem{proposition}{Proposition}
\newtheorem{corollary}{Corollary}

\newtheorem{lemma}{Lemma}

\begin{document}

\title{Low-Complexity Optimal Scheduling over Time-Correlated Fading Channels with\\ ARQ Feedback}

\author{\emph{Wenzhuo Ouyang, Atilla Eryilmaz, and Ness B. Shroff}
\vspace{-17pt}

\thanks{Wenzhuo Ouyang is with the Department of ECE, Rice University (e-mail: wenzhuo.ouyang@rice.edu). Atilla Eryilmaz is with the Department of ECE, The Ohio State University (e-mail: eryilmaz@ece.osu.edu).
Ness B. Shroff holds a joint appointment in both the Department of ECE and the Department of CSE at The Ohio State University (e-mail: shroff@ece.osu.edu).}

\thanks{A preliminary version of this paper appeared in WiOpt 2012 \cite{WiOpt_version}.}

\thanks{This work is partly supported by NSF grants CNS-0721434, CNS-0831919, CNS-0953515, CCF-0916664, DTRA grant HDTRA 1-08-1-0016, Army Research Office MURI Awards W911NF-08-1-0238 and W911NF-07-1-0376.}
}
\maketitle

\begin{abstract}

We investigate the downlink scheduling problem under Markovian ON/OFF fading channels, where the instantaneous channel state information is not directly accessible, but is revealed via ARQ-type feedback. The scheduler can exploit the temporal correlation/channel memory inherent in the Markovian channels to improve network performance. However, designing low-complexity and throughput-optimal algorithms under temporal correlation is a challenging problem. In this paper, we find that under an average number of transmissions constraint, a low-complexity index policy is throughput-optimal. The policy uses Whittle's index value, which was previously used to capture opportunistic scheduling under temporally correlated channels. Our results build on the interesting finding that, under the intricate queue length and channel memory evolutions, the importance of scheduling a user is captured by a simple multiplication of its queue length and Whittle's index value. The proposed queue-based index policy has provably low complexity. Numerical results show that significant throughput gains can be realized by exploiting the channel memory using the proposed low-complexity policy.
\end{abstract}\vspace{-10pt}

\section{Introduction}

In wireless networks with randomly fluctuating channels, intelligently scheduling users is critical for achieving high network efficiency. Under the assumption that the scheduler possesses accurate instantaneous Channel State Information (CSI),
many sophisticated scheduling algorithms have been proposed and extensively studied (e.g., \cite{backpressure}-\cite{Eryilmaz05}).

In practice, accurate instantaneous CSI is difficult to obtain at the scheduler. Hence, in this work we consider the important scenario where the instantaneous CSI is not directly accessible to the scheduler, but is instead revealed through ARQ-type feedback only \emph{after} each scheduled data transmission. Many works have focused on scheduling algorithms design with imperfect CSI, where the channel state is considered independent and identically distributed (i.i.d.) processes across time (e.g., \cite{Allerton}-\cite{Wenzhuo_wiopt}). On the other hand, although the i.i.d. channel model facilitates more tractable analysis, it does not capture the time-correlation of the fading channels. ARQ-based protocols over time-correlated channels are studied in \cite{CHARQ}-\cite{Green_HARQ} under the scenarios where user scheduling is not required.

The time-correlation or channel memory inherent in the fading channels can be exploited by the scheduler for more informed decisions, and hence to obtain large throughput/utility gains (e.g., \cite{Infocom11}-\cite{CelicModiano}). Under imperfect CSI, channel memory, and limited network resources, designing efficient scheduling schemes is highly challenging. This is because the scheduler needs to optimally balance the intricate `exploitation-exploration tradeoff', i.e., to decide whether to exploit the channels with more up-to-date CSI, or to explore the channels with outdated CSI.

In this work, we study downlink scheduling with imperfect CSI and time correlated channels where, differing from works \cite{Infocom11}-\cite{Zhao_index} in this domain, the packets destined to each user randomly arrive in time, and are stored in a corresponding observable data queue before transmission. As a result, the queue lengths randomly evolve with time. Our goal is to design scheduling algorithm that is throughput optimal, i.e., no scheduling policy can ensure system stability for arrival rates that are not supportable by the proposed scheduler. Considering queue lengths along with imperfect CSI and time correlation is highly challenging because to develop throughput-optimal scheduler requires a complex characterization of the interplay between user scheduling, channel memory evolution and queue evolution. Traditional techniques, which assume known service rate (e.g.\cite{Glazebrook}\cite{Glazebrook2}), or assume i.i.d. channel state process and are based on minimizing instantaneous Lyapunov drift in each slot (e.g., scheduling user with maximal instantaneous product of queue length and transmission rate \cite{backpressure}\cite{MWM}), does not apply in this context.

Under this model, because of the aforementioned complications, traditional Dynamic Programming based approaches can be used for designing scheduling schemes, but are intractable due to the well-known `curse of dimensionality'. In \cite{Neely_RR}\cite{Neely_RR2}, simple round-robin based scheduling policies are shown to possess the throughput-optimality property. The optimality of greedy scheduling algorithm are proven in \cite{ZhaoTWC}\cite{SM_IT}. However, these schemes \cite{Neely_RR}-\cite{SM_IT} are only optimal in the regime where users have \emph{identical} ON/OFF Markovian channel statistics. In \cite{KrishnaModiano}\cite{CelicModiano}, throughput-optimal frame-based policies are proposed. These policies rely on solving a Linear Programming in each frame, which is hindered by the curse of dimensionality where the computational complexity grows exponentially with the network size.

In this work, we study throughput-optimal downlink scheduling under imperfect CSI over heterogeneous Markovian fading channels. We consider time-correlation by modeling the fading channel as an `ON/OFF' Markov chain. Differing from the previous works \cite{Neely_RR}-\cite{CelicModiano} that consider scheduling problems under strict interference constraints (e.g., only one user can be scheduled at each time slot), we assume that each user occupies a dedicated channel, i.e., all users can transmit simultaneously, but the \emph{long-term average} number of transmissions is limited. In this setup, we show that a low complexity scheduling policy is throughput optimal. Such a constraint on long-term average number of transmissions can be used to limit the long-term energy consumption. An example to limit the energy consumption is the \emph{green cellular networks} (e.g., \cite{DegFreedom}-\cite{SonKrishnamachari}). It is estimated that the cellular base stations consume $4.5$ GW of power globally, which corresponds to more than $40$ million metric tons of CO$_2$ emission and over $\mathdollar 10$ billion electricity bill annually \cite{DegFreedom}\cite{KrishnamachariLiu}. With energy expenditure rising by $15$-$20 \%$ each year, an important objective in green cellular networks design is to reduce the long-run average number of data transmissions to decrease energy consumption \cite{KrishnamachariLiu}. Therefore, it is of great interest to understand the relationship between the achievable throughput region and the constraint on the long-term average number of transmissions. The results proposed in this work can be applied to green cellular networks for throughput-optimal scheduling under imperfect CSI and the long-term average energy constraint.

Our contributions are as follows:

\noindent$\bullet$  Under the constraint on the long-term average number of transmissions, we propose a low-complexity \emph{throughput-optimal} policy. The policy operates over separate time frames and, in each time frame, tries to maximize a queue-weighted average sum-throughput. We are able to conduct a \emph{frame-based Lyapunov analysis} to this policy and prove its optimality by showing that it minimizes the average Lyapunov drift over each frame. Compared to the traditional approaches for i.i.d. channels based on minimizing \emph{instantaneous} Lyapunov drift each slot, the frame-based approach is useful for analysis in scenarios with time-correlated channels. The per-frame computational complexity is at most $O((2\tau+1)N \log (2\tau+1)N)$ with the number of users $N$, where $\tau$ is a control parameter independent of $N$. Therefore, the policy does not suffer from the curse of dimensionality.

\noindent$\bullet$ The proposed policy builds on Whittle's index analysis of Restless Multi-armed Bandit Problem (RMBP) \cite{Whittle}, where Whittle's index value is used to measure the importance of scheduling a user under the time-correlated channel \cite{Wenzhuo_infocom12}. Whittle's index policies are known to have optimality properties in various RMBP processes and
have been shown to have low-complexity (e.g., \cite{Oppo}\cite{Glazebrook}\cite{Glazebrook2}). We find that, interestingly, under the coupled queue length and channel memory evolution, the importance of scheduling a user is measured by a simple \emph{multiplication} of the queue length and Whittle's index value that is given in closed-form. This property is essential for the low-complexity nature of our policy.

\begin{figure}
\centering
\includegraphics[width=2.7in]{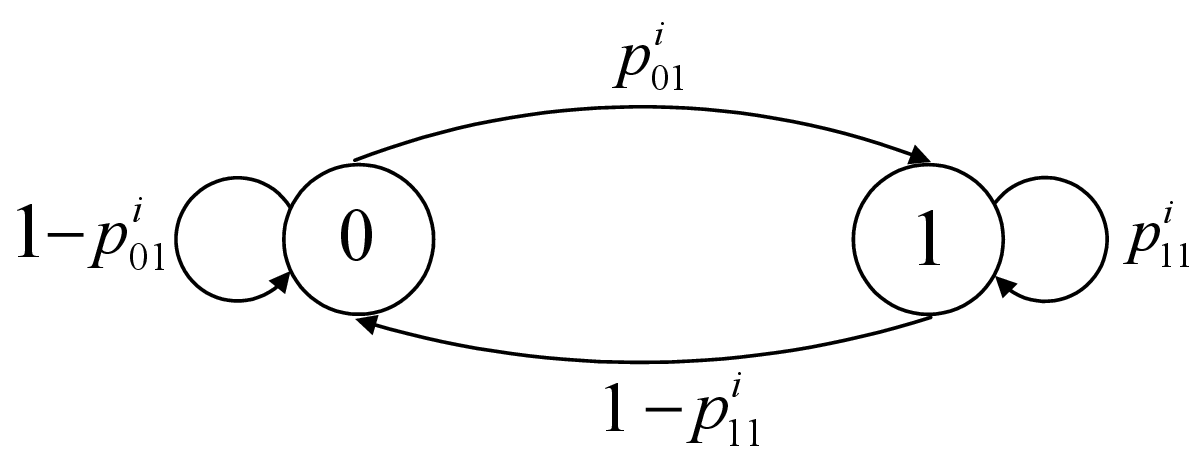}
\vspace{-2pt}
\caption{Two state Markov Chain model.}
\label{fig:chain}\vspace{-8pt}
\end{figure}

\section{System Model}

\subsection{Downlink Scheduling Problem}

We consider a time-slotted wireless downlink network with one base station and $N$ users, where each user $i$ occupies a dedicated wireless channel. The channel state of user $i$, denoted by $C_i[t]$ at slot $t$, evolves according to an ON/OFF Markov chain across time slots within the state space $\mathcal{S}=\{0, 1\}$, independently across channels. When the channel is in state `1', one packet can be successfully transmitted, otherwise no packet can be delivered. As shown in Fig.~\ref{fig:chain}, the channel state evolution is represented by the transition probabilities
\begin{align}p^i_{11}:=& \Pr\big(C_i[t]{=}1 \big | C_i[t{-}1]{=}1\big),\nonumber\\
p^i_{01}:=& \Pr\big(C_i[t]{=}1 \big | C_i[t{-}1]{=}0 \big).\nonumber
\end{align}

We assume that the Markovian channels are positively correlated, i.e., $p^i_{11} > p^i_{01}$ for $i{=}1,2,\cdots, N$. This assumption is commonly made in this field (e.g., \cite{Wenzhuo_infocom12}\cite{Neely_RR}\cite{KrishnaModiano}\cite{sugu_aslm}), which means that auto-correlation of the channel state process is non-negative \cite{Javidi}. This means, roughly speaking, that the Markov channel is more likely to stay in its state than changing to another state, which captures the typical slow fading or fast transmission scenarios. For ease of presentation, we ignore the trivial case when $p^i_{11}=1$ or $p^i_{01}=0$, $i\in\{1,\cdots, N\}$.

At the beginning of each time slot, the scheduler chooses users for data transmission. The scheduling decisions are made without the exact knowledge of the channel state in the current slot. Instead, the accurate ON/OFF channel state of a scheduled user is revealed via ACK/NACK feedback from the receiver, only at the end of each slot following data transmission.

We consider the class $\Phi$ of (possibly non-stationary) scheduling policies that make scheduling decisions based on the history of observed channel states, arrival processes, and scheduling decisions. Under the aforementioned restrictions on average energy consumption, the scheduling schemes are subject to the constraint that the long-term average number of scheduled transmissions is under $M$,

\vspace{-12pt}\begin{align}
\limsup_{T\rightarrow \infty} \frac{1}{T}\mathbb{E}\Big[\sum_{t=0}^{T-1} \sum_{i=1}^{N} a_i^{\phi}[t]\Big ]\leq M, \label{eq:constraint}
\end{align}

\noindent where $a_i^{\phi}[t]\in\{0,1\}$ indicates whether user $i$ is scheduled at slot $t$ under policy $\phi \in \Phi$, and $M\leq N$.

Data packets destined for different users are stored in separate queues before transmission. The queue length for user $i$ is denoted by $q_i[t]$ at slot $t$. We assume that the packet arrivals for the $i$-th user form an \emph{i.i.d.} process $A_i[t]$ with mean $\lambda_i$ and a bounded second moment. Hence, the $i$-th data queue evolves as $q_i[t{+}1]{=}\max \{ 0, q_i[t] {-}a_i[t]{\cdot} C_i[t] \}{+} A_i[t]$.

\begin{figure}
\centering
\includegraphics[width=3in]{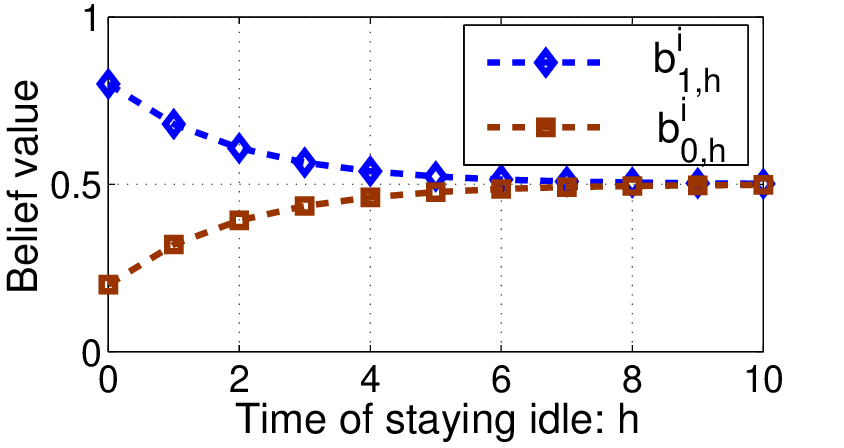}
\vspace{-5pt}
\caption{Belief value evolution, $p^i_{11}=0.8$, $p^i_{01}=0.2$, $b^i_s=0.5$.}
\label{fig:belief_evol}\vspace{-18pt}
\end{figure}

\vspace{-6pt}\subsection{Belief Value Evolution}

The scheduler maintains a belief value $\pi_i[t]$ for each channel $i$, defined as the probability of channel $i$ being in state $1$ at the beginning of $t$-th slot conditioned on the past channel state observations. The belief values are hence updated according to the scheduling decisions and accurate channel state feedbacks,
\begin{align}
\pi_i[t+1]=
\begin{cases}
p^i_{11}& \text{if $a_i[t]=1$ and $C_i[t]=1$,}\\
p^i_{01}& \text{if $a_i[t]=1$ and $C_i[t]=0$,} \\
Q_i(\pi_i[t])& \text{if $a_i[t]=0$,}
\end{cases} \label{eq:evolve}
\end{align}
where $Q_i(x){=}x p^i_{11}+(1{-}x)p^i_{01}$ is the belief evolution operator when user $i$ is not scheduled in the current slot. In our setup, the belief values are known to be sufficient statistics to represent the past scheduling decisions and channel state feedback \cite{Sondik_thesis}. In the meanwhile, the belief value $\pi_i[t]$ is the expected throughput for user $i$ if it is scheduled in slot $t$.

For the $i$-th user, we use $b^i_{c,h}$ to denote the state of its belief value when the most recent channel state was observed $h$ time slots ago and was in state $c\in \{0,1\}$. The closed form expression of $b^i_{c,h}$ can be calculated from (\ref{eq:evolve}) and is given as
\begin{align}
b^i_{0,h}{=}\frac{p^i_{01}\hspace{1pt}{-}\hspace{1pt}(p^i_{11}{-}p^i_{01})^h
p^i_{01}}{1+p^i_{01}-p^i_{11}},
b^i_{1,h}{=}\frac{p^i_{01}{+}(1{-}p^i_{11})(p^i_{11}{-}p^i_{01})^h}{1+p^i_{01}-p^i_{11}}.
\nonumber
\end{align}

As depicted in Fig.~\ref{fig:belief_evol}, if the scheduler is never informed of the $i$-th user's channel state, the belief value monotonically converges to the stationary probability $b_s^{i}{:=}{p^i_{01}}/{(1+p^i_{01}-p^i_{11})}$ of the channel being in state $1$. We assume that the belief values of all channels are initially set to their stationary values. It is then clear that, based on (\ref{eq:evolve}), each belief value $\pi_i[t]$ evolves over a countable state space, denoted by $\mathcal{B}_i{=}\{b^i_{s}, b^i_{c,h}: c \hspace{1pt}{\in}\hspace{1pt} \{0,1\}, h\hspace{1pt}{\in}\hspace{1pt} \mathbb{Z}^+ \}$.

\vspace{-10pt}\subsection{Network Stability Region and Achievable Rate Region}

We adopt the following definition of queue stability
\cite{MWM}: queue $i$ is stable if there exists a limiting
stationary distribution $F_i$ such that $\lim_{t\rightarrow \infty}
P( q_i[t] \leq q)= F_i(q)$.  The \emph{network stability region} $\bm \Lambda$ is defined as the closure of the set of arrival rate vectors supported by all policies in class $\Phi$ that does not lead to system instability while abiding by the constraint (\ref{eq:constraint}). A policy is called \emph{throughput optimal} if, for any arrival rate vector $\bm \lambda$ within arbitrary $\epsilon$ interior of $\bm \Lambda$, i.e., $\bm\lambda+\epsilon\bm 1 \in\bm\Lambda$, all queues are stable under the policy and constraint (\ref{eq:constraint}) is satisfied.

In the meanwhile, we define the \emph{achievable rate region} $\bm \Gamma$ as the closure of the set of service rate vectors $\mathbf{\bm \gamma}$ that can be achieved by all policies, i.e.,
\vspace{-6pt}\begin{align}
\bm \Gamma{=} Cl\big\{ \mathbf{\bm \gamma}: &\exists \phi \in \Phi
\text{\hspace{2pt}with\hspace{2pt}} \gamma_i{=}\liminf_{T\rightarrow \infty} \frac{1}{T} \mathbb{E}\big[\sum_{t=0}^{T-1} \pi_i[t] \cdot a_i^{\phi}[t]\big], \nonumber\\
&i=1, \cdots, N,\ \text{subject to constraint} \ (\ref{eq:constraint})\big\}, \label{eq:rate_reg}
\end{align}
where $Cl\{\cdot\}$ denotes the closure of the set. The rate region is convex since randomization can be performed among different policies. The achievable rate region $\bm \Gamma$ contains the set of the expected service rate vectors that can be achieved with all the policies in $\Phi$,  in the system with \emph{infinitely backlogged queues}.

\vspace{-8pt}\section{Optimal Policy for Weighted Sum-throughput Maximization}
\label{sec:sum_thr}

In this section, we postpone discussion on queue evolution and consider a simplified problem with infinitely backlogged queues, and derive the corresponding optimal policy for weighted sum-throughput maximization. The policy introduced here, which is based on scaling the Whittle's index values, is useful to characterize the boundary point of the achievable rate region $\bm \Gamma$, and is also an important part in the throughput-optimal policy in the next section that stabilizes all arrival rates within the system stability region $\bm\Lambda$ -- the main result of the paper.

\vspace{-9pt}\subsection{Weighted Sum-throughput Maximization Problem}
\label{sec:sum_thr_non_trun}

Consider the following weighted sum-throughput maximization problem $\Psi(\mathbf{r}, M)$ for a given vector $\mathbf{r}=(r_i)_{i=1}^N$, where the expected service rate for each user $i$ is scaled by a non-negative factor $r_i$,
\vspace{-5pt}\begin{align}
\max_{\phi \in \Phi}& \ \liminf_{T\rightarrow \infty} \frac{1}{T} \mathbb{E}\Big[\sum_{t=0}^{T-1} \sum_{i=1}^N r_i {\cdot} \pi_i[t] {\cdot} a_i^{\phi}[t] \Big]\label{eq:obj_rel}\\
\text{s.t.}&\hspace{5pt} \limsup_{T\rightarrow \infty} \frac{1}{T}\mathbb{E}\Big[\sum_{t=0}^{T-1} \sum_{i=1}^{N} a_i^{\phi}[t]\Big ]\leq M. \label{eq:cons_rel}
\end{align}

The above problem $\Psi(\mathbf{r}, M)$ is hence a constrained Partially Observable Markov Decision Process (CPOMDP) \cite{Eitan}\cite{Meyn_CPOMDP}.

\vspace{-9pt}\subsection{Whittle's Index for Restless Multi-armed Bandit Problem}

The problem~(\ref{eq:obj_rel})-(\ref{eq:cons_rel}) appears difficult because of the complex `exploitation - exploration' tradeoff. To tackle this problem, we study it in the framework of the Restless Multiarmed Bandit Problem (RMBP) \cite{Whittle} and make use of the associated Whittle's indexability analysis. We next give a brief review of the Whittle's indices for RMBP.

RMBPs refer to a collection of sequential dynamic resource allocation problems where several independently evolving projects compete for service. In each slot, a subset of these competing projects is served. The state of each project stochastically evolves over time, based on the current state of the project and on whether the project is served in the slot. Serving a project brings a reward whose value depends on its state. Hence, in RMBPs, the controller needs to consider the fundamental tradeoff between decisions that bring high instantaneous rewards, versus those decisions that bring better future rewards but sacrifices the instantaneous rewards. Solving RMBPs are known to be PSPACE-hard \cite{Tsitsiklis} in general.

Whittle's index analysis \cite{Whittle} for RMBPs considers the following \emph{virtual system}: in each slot, the controller makes one of the two decisions for each project $P$: (1) Serve project $P$ and accrue an immediate reward as a function of its state which is the same as in the original RMBP. (2) Do not serve project $P$ and obtain an immediate reward $\omega$ for passivity. The state evolution of the project $P$ is the same as in the original RMBP, depending on its current state and current action. In this virtual system, the design goal is to maximize the long-term expected reward by balancing the `reward for serving' and the `subsidy for passivity' in each slot.

Letting $\mathcal{I}(\omega)$ denote the set of states of project $P$ in which the optimal action is to stay passive, the Whittle's indexability condition is defined as follows.

\textit{Project $P$ is Whittle indexable if the set $\mathcal{I}(\omega)$ monotonically increases from $\emptyset$ to the state space $\mathcal{S}$ of project $P$, as $\omega$ increases from $-\infty$ to $\infty$. The RMBP is Whittle indexable if \textit{every} project is Whittle indexable.}

\vspace{5pt} If Indexability holds, for each state $s$ of a project, the \emph{Whittle's index} $W(s)$ is defined as the infimum of $\omega$ in which it is optimal to stay idle in the $\omega$-subsidized system, i.e.,
\vspace{-4pt}\begin{align}
W(s)=\inf\{ \omega: s \in \mathcal{I}(\omega) \}. \nonumber
\end{align}

\vspace{-5pt}Under an average constraint on the number of projects scheduled per slot, it is known that, upon the satisfaction of the Indexability condition, an optimal algorithm exists based on the `Whittle's indices': activate the projects with large Whittle's index value \cite{Whittle}.

The RMBP theories and the associated Whittle's indices can be used in our downlink scheduling problem. Here, each downlink user corresponds to a project in the RMBP, with the associated state being the belief value of its channel. Correspondingly, the project is considered served if the user is scheduled for data transmission at a slot. Hence the Whittle's index policy is very attractive to provide optimal solutions to our problem, as we shall elaborate in the rest of the paper.

\vspace{-8pt}\subsection{Optimal Policy for Weighted Sum-throughput Maximization}

It was shown in that our downlink scheduling problem is Whittle indexable \cite{Zhao_index}, and, under uniform weight vector $\mathbf{r}{=}\mathbf{1}$, an optimal policy for problem $\Psi(\mathbf{1}, M)$ exists based on Whittle's indexability analysis of Restless Multi-armed Bandit Problem \cite{Wenzhuo_infocom12}. Specifically, for channel $i$, a closed form \emph{Whittle's index value} $W_i^{\mathbf{1}}(\pi)$ is assigned to each belief state $\pi \in \mathcal{B}_i$. These indices intelligently capture the exploitation-exploration value to be gained from scheduling the user at the corresponding belief state \cite{Wenzhuo_infocom12}. The closed form expression of the Whittle's index value $W_i^{\mathbf{1}}(\pi), \pi \in \mathcal{B}_i$, is given as follows \cite{Wenzhuo_infocom12}\cite{Zhao_index},
\begin{align}
\label{eq:indices}
W_i^{\mathbf{1}}(\pi)
{=}\hspace{-3pt}\begin{cases}
\frac{(\pi-Q_i(\pi)) (h+1)+Q_i(\pi)}{1-p^i_{11}+(\pi-Q_i(\pi))h+Q_i(\pi)} &\text{if $p^i_{01} {\leq} \pi{=}b^i_{0,h} {<} b^i_s$} \\
\frac{p^i_{01}}{(1-p^i_{11})(1+p^i_{01}-p^i_{11})+p^i_{11}} &\text{if $b^i_s \leq \pi\leq p^i_{11}$}
\end{cases}
\end{align}

It was shown that $W_i^{\mathbf{1}}(\pi)$ monotonically increases with $\pi$ and satisfies $W_i^{\mathbf{1}}(\pi) \in [0,1]$ \cite{Wenzhuo_infocom12}\cite{Zhao_index}. In the following lemma, we give the optimal algorithm to the problem $\Psi({\mathbf{r}}, M)$ with arbitrary non-negative weight vector $\mathbf{r}$. The proof of the lemma follows the line of \cite{Whittle} and is re-proven in Appendix~\ref{appen:threshold}.

\begin{lemma}
\label{lemma:thres_index}
There exists an optimal stationary policy $\phi^*(\mathbf{r}, M)$ for problem $\Psi(\mathbf{r}, M)$ (cf. (\ref{eq:obj_rel})-(\ref{eq:cons_rel})), parameterized by a user index $i^*$, a threshold $\omega^*$ and a randomization factor $\rho^*$, such that

\noindent(i) The scheduler maintains an $\mathbf{r}$-weighted index value $W^{\mathbf{r}}_i(\pi_i[t])=r_i \cdot W_i^{\mathbf{1}}(\pi_i[t])$ for user $i$.
\vspace{5pt}

\noindent(ii) User $i$ is scheduled if $W^{\mathbf{r}}_i(\pi_i[t])\hspace{1pt}{>}\hspace{1pt}\omega^*$, or if $W^{\mathbf{r}}_i(\pi_i[t])\hspace{1pt}{=}\hspace{1pt}\omega^*$ with $i{>}i^*$. User $i$ stays idle if \hspace{3pt}$W^{\mathbf{r}}_i(\pi_i[t])\hspace{1pt}{<}\hspace{1pt}\omega^*$, or if $W^{\mathbf{r}}_i(\pi_i[t])\hspace{1pt}{=}\hspace{1pt}\omega^*$ with $i{<}i^*$.
If $W^{\mathbf{r}}_i(\pi_i[t])\hspace{1pt}{=}\hspace{1pt}\omega^*$ with $i= i^*$, user $i$ is scheduled with probability $\rho^*$.
\vspace{5pt}

\noindent(iii) The parameters $i^*$, $\omega^*$ and $\rho^*$ are such that the long-term average number of transmissions equals $M$.
\end{lemma}

\noindent \textbf{Remarks:} Interestingly, by multiplying the Whittle's index values $W_i^{\mathbf{1}}(\pi_i[t])$ with $r_i$, the optimal policy $\phi^*(\mathbf{1}, M)$ extends to more general problem $\Psi({\mathbf{r}}, M)$. This property is important for designing the throughput-optimal policy in Section~\ref{sec:QWI_policy}.

\vspace{-10pt}\subsection{Approximate $i^*$, $\omega^*$ and $\rho^*$ using State Space Truncation}
\label{sec:thr_approx}

Note that the parameters $i^*$, $\omega^*$ and $\rho^*$ need to be carefully chosen to satisfy the complementary slackness condition, i.e., Lemma 1(iii). While directly finding these parameters may be difficult, we next introduce an algorithm to derive approximate values of $i^*$, $\omega^*$ and $\rho^*$ based on a fictitious model over \emph{truncated belief state space}. This fictitious model facilitates more tractable design and analysis. More importantly, we shall show that, when implementing these approximate values over the original untruncated system, the performance will get arbitrary close to the optimality.

Recall that the belief value $\pi_i[t]$ evolves over a countable state space $\mathcal{B}_i$ for user $i$ and approaches the stationary value if the channel is not active for a long time. This motivates us to consider the following fictitious belief evolution model over the truncated state space: the belief value of a user is set to its steady state (i.e., its channel state history is entirely forgotten) if the corresponding channel has not been scheduled for a long time, say $\tau$ slots. We use $\pi_i^{\tau}[t]$ to denote this `heuristic belief value'. The evolution of $\pi_i^{\tau}[t]$ is hence,
\begin{align}
&\pi_i^{\tau}[t+1]=\nonumber\\
&\begin{cases}
p^i_{11}& \text{if $a_i[t]=1$ and $C_i[t]=1$,}\\
p^i_{01}& \text{if $a_i[t]=1$ and $C_i[t]=0$,} \\
Q_i(\pi_i[t])& \text{if $a_i[t]=0$, $\prod_{k=1}^{\tau-1}{\big(1{-}a_i[t{-}k]\big)}=0$,}\\
b^i_{s} & \text{if $\prod_{k=0}^{\tau-1}{\big(1{-}a_i[t{-}k]\big)}=1$.}
\end{cases}\label{eq:belief_evol_trun}
\end{align}

We let $\mathcal{B}_i^{\tau}$ denote the truncated state space for the $i$-th user, i.e., $\mathcal{B}_i^{\tau}{=}\{b^i_{s}, b^i_{c,l}: c \hspace{1pt}{\in}\hspace{1pt} \{0,1\}, l\hspace{1pt}{=}\hspace{1pt} 1,2,\cdots,\tau \}$ and let $\mathbf{B}^{\tau}=[\mathcal{B}_1^{\tau},\cdots,\mathcal{B}_N^{\tau}]$. Over the fictitious truncated state space, we consider the following policy $\phi^{trunc}_{j,\omega,\rho}$:

\vspace{8pt}\textbf{Policy ${\phi^{trunc}_{j,\omega,\rho}}$ over the truncated state space:} \emph{User $i$ is scheduled if $W^{\mathbf{r}}_i(\pi_i^{\tau}[t])\hspace{1pt}{>}\hspace{1pt}\omega$, or if $W^{\mathbf{r}}_i(\pi_i^{\tau}[t])\hspace{1pt}{=}\hspace{1pt}\omega^*$ with $i{>}j$. User $i$ stays idle if\hspace{3pt}$W^{\mathbf{r}}_i(\pi_i^{\tau}[t])\hspace{1pt}{<}\hspace{1pt}\omega$, or if $W^{\mathbf{r}}_i(\pi_i^{\tau}[t])\hspace{1pt}{=}\hspace{1pt}\omega^*$ with $i{<}j$.
If \hspace{1pt} $W^{\mathbf{r}}_i(\pi_i^{\tau}[t])\hspace{1pt}{=}\hspace{1pt}\omega$ with $i{=}j$, it is scheduled with probability $\rho$.}


\vspace{8pt}Under this setup, we let the parameter ${\alpha}^{\tau}_i(j, \omega, \rho)$ denote the long-term expected fraction of time transmitting to user $i$, i.e.,
\begin{align}
\label{eq:alpha_tau_def}
{\alpha}^{\tau}_i(j, \omega, \rho)=\limsup_{T\rightarrow \infty} \frac{1}{T} \mathbb{E}\Big[ \sum_{t=0}^{T-1} a_i^{\phi^{trunc}_{j, \omega, \rho}}[t]\Big],
\end{align}
\vspace{-2pt}where $a_i^{\phi^{trunc}_{j, \omega, \rho}}[t]\in\{0,1\}$ indicates whether user $i$ is scheduled at time $t$ under policy $\phi^{trunc}_{j, \omega, \rho}$. The closed-form expression of ${\alpha}^{\tau}_i(j, \omega, \rho)$ is given by the following lemma. The proof of the lemma is given in Appendix~\ref{sec:alpha_tau_proof}.

\vspace{-3pt}\begin{lemma}
\label{lemma:alpha_tau}
Let the value $\tau_0$ be
{\small\begin{align}\label{eq:tau0}
\hspace{-9pt}\tau_0{=} \Big\lceil4 \max\big\{\frac{1}{{-}\log(p^i_{11}{-}p^i_{01})}, \frac{1}{\log^2(p^i_{11}{-}p^i_{01})}, i{=}1,{\cdots}, N\big\}\Big\rceil.
\end{align}}

Over the truncated state space and under policy $\phi^{trunc}_{j, \omega, \rho}$, if $\tau>\tau_0$, the following hold for ${\alpha}^{\tau}_i(j, \omega, \rho)$,

\vspace{2pt}\noindent(i) The closed-form expression of ${\alpha}^{\tau}_j(j, \omega, \rho)$ is given by
\begin{align}
&{\alpha}^{\tau}_j(j, \omega, \rho)\nonumber\\
={\hspace{-5pt}}&\begin{cases}
\frac{\rho(b^j_{0,h}-b^j_{0,h+1})+1-p^j_{11}+b^j_{0,h{+}1}}{\rho b^j_{0,h}{+}(1{-}\rho)b^j_{0,h\hspace{-1pt}{+}\hspace{-1pt}1}{+}(1\hspace{-1pt}{-}\hspace{-1pt}p^j_{11})(h{+}1{-}\rho)} &\text{if $\omega{=}W^{\mathbf{r}}_j(b^j_{0,h})$, $h{<}\tau$}\\
\frac{\rho(b^j_{0,\tau}-b^j_{s})+1-p^j_{11}+b^j_{s}}{\rho b^j_{0,\tau}+(1-\rho)b^j_{s}+(1-p^j_{11})(\tau+1-\rho)} &\text{if $\omega{=}W^{\mathbf{r}}_j(b^j_{0,\tau})$} \\
\frac{\rho(1-p^j_{11}+b^j_{s})}{(1+\tau\rho)(1-p^j_{11})+\rho b^j_{s}} &\text{if $\omega{=}W^{\mathbf{r}}_j(b^j_s)$}\\
0 &\text{if $\omega{>}W^{\mathbf{r}}_j(b^j_s)$.}
\nonumber
\end{cases}
\end{align}
The closed-form expression of ${\alpha}^{\tau}_i(j, \omega, \rho), i\neq j$ is given by
\begin{align}
&{\alpha}^{\tau}_i(j, \omega, \rho)\nonumber\\
={\hspace{-5pt}}&\begin{cases}
\frac{1-p^i_{11}+b^i_{0,h{+}1}}{b^i_{0,h{+}1}{+}(1{-}p^i_{11})(h{+}1)}&\text{if $h{<}\tau$, $\omega{=}W^{\mathbf{r}}_i(b^i_{0,h})$, $i{<}j$}\\
\frac{1-p^i_{11}+b^i_{0,h}}{b^i_{0,h}{+}(1{-}p^i_{11})h}&\text{if $h{\leq}\tau$, $\omega{=}W^{\mathbf{r}}_i(b^i_{0,h})$, $i>j$;}\\
&\text{or if $h{\leq}\tau$,$W^{\mathbf{r}}_i(b^i_{0,h{-}1}){<}\omega{<}W^{\mathbf{r}}_i(b^i_{0,h})$}\\
\frac{1-p^i_{11}+b^i_{s}}{b^i_{s}+(1-p^i_{11})(\tau+1)} &\text{if $\omega{=}W^{\mathbf{r}}_i(b^i_{0,\tau})$, $i<j$;}\\
&\text{or if $\omega{=}W^{\mathbf{r}}_i(b^i_s)$, $i>j$}\\
0 &\text{if $\omega{=}W^{\mathbf{r}}_i(b^i_s)$, $i<j$;}\\
&\text{or if $\omega{>}W^{\mathbf{r}}_i(b^i_s)$.}
\nonumber
\end{cases}
\end{align}

\vspace{-3pt}\noindent(ii) For fixed $\pi_j{\in}\{b^j_{0,1},b^j_{0,2},{\cdots},b^j_{0,\tau},b^j_{s}\}$, ${\alpha}^{\tau}_j(j, W_j^{\mathbf{r}}(\pi_j),\rho)$ strictly increases with $\rho$. For fixed $\rho$, ${\alpha}^{\tau}_i(j,W_i^{\mathbf{r}}(\pi_i),\rho)$ strictly decreases with $\pi_i$ for $\pi_i\in\{b^i_{0,1},b^i_{0,2},\cdots,b^i_{0,\tau},b^i_{s}\}$ and all $i$.
\end{lemma}

We approximate the optimal values $i^*$, $\omega^*$ and $\rho^*$ (defined in Lemma~\ref{lemma:thres_index}) using the fictitious truncated state space model. The approximate value $i_{\tau}$, $\omega_{\tau}$ and $\rho_{\tau}$ are such that, under policy $\phi^{trunc}_{i_{\tau}, \omega_{\tau}, \rho_{\tau}}$ over the truncated state space, the long-term average number of transmissions equals $M$, i.e.,
\vspace{-4pt}\begin{align}
\sum_{i=1}^N {\alpha}^{\tau}_i(i_{\tau}, \omega_{\tau}, \rho_{\tau})=M\label{eq:tau_constr}.
\end{align}

\vspace{-5pt}Note that, equation~(\ref{eq:tau_constr}) is the truncated-state-space correspondence of Lemma~\ref{lemma:thres_index}(iii). We next design an algorithm, denoted by $G^{\tau}(\mathbf{r},M)$, to calculate $i_{\tau}$, $\omega_{\tau}$ and $\rho_{\tau}$, described to the right and explained next.

\renewcommand{\thealgorithm}{}

\algnewcommand{\algorithmicgoto}{\textbf{go to}}%
\algnewcommand{\Goto}[1]{\algorithmicgoto~\ref{#1}}%

\begin{algorithm}
  \caption{\hspace{-6pt}$G^{\tau}(\mathbf{r},M)$\textbf{:} Calculation of $i_{\tau}$, $\omega_{\tau}$ and $\rho_{\tau}$}\label{alg:thres}
  \begin{algorithmic}[1]
  \State $\text{TxTime}[i]=1 \text{ for all $i\in\{1,\cdots, N\}$}$
  \State $\text{TotalTime}=N$
  \State \text{\textbf{struct} Index}
  \State $\{$\text{ float value}
  \State \text{ \hspace{3pt} int user}
  \State $\}$ $\mathbf{I}[(2\tau+1)N],\mathbf{w}[(2\tau+1)N]$
  \vspace{5pt}\State $j=0$
  \For{$i=1$ \text{to} $N$}
  \For{each $\pi_i \in \mathcal{B}_i^{\tau}$}
        \State $W_i^{\mathbf{r}}(\pi_i)= r_i\cdot W_i^{\mathbf{1}}(\pi_i)$
    \State $\mathbf{I}[j].$\text{value}$=W_i^{\mathbf{r}}(\pi_i)$
    \State $\mathbf{I}[j].$\text{user}$=i$
    \State $j\gets j+1$
  \EndFor
  \EndFor
  \vspace{5pt}\State $\mathbf{w}=$\text{sort}$(\mathbf{I})$ \Comment{Sort the elements in $\mathbf{I}$ in increasing order }
  \Statex \text{\hspace{0.8in}of the index value and outputs to vector $\mathbf{w}$.}
  \Statex \text{\hspace{0.8in}For index values that are equal, they are or-}
  \Statex \text{\hspace{0.8in}dered in increasing order of the associated}
  \Statex \text{\hspace{0.8in}user index.}
    \For{$k=1$ \text{to} \text{size}$(\mathbf{w})$}
    \vspace{5pt}\State $\text{NewTime}[\mathbf{w}[k].\text{user}]={\alpha}^{\tau}_{\mathbf{w}[k].\text{user}}(\mathbf{w}[k].\text{value}, 1)$
\State  $\text{TimeDiff}=\text{TxTime}[\mathbf{w}[k].\text{user}]{-}\text{NewTime}[\mathbf{w}[k].\text{user}]$
    \State $\text{TotalTime}=\text{TotalTime}-\text{TimeDiff}$
    \If{$\text{TotalTime}<M$}
    \State $i_{\tau}=\mathbf{w}[k-1].\text{user}$
    \State $\omega_{\tau}=\mathbf{w}[k{-}1].\text{value}$
    \State $\text{TxTime}[\mathbf{w}[k{-}1].\text{user}]=M{-}\hspace{-9pt}\sum\limits_{i\neq \mathbf{w}[k{-}1].\text{user}}\hspace{-9pt}\text{TxTime}[i]$
    \State $\rho_{\tau}=\beta_{\mathbf{w}[k{-}1].\text{user}}(\omega_{\tau},\text{TxTime}[\mathbf{w}[k{-}1].\text{user}])$
    \State \text{\textbf{Break}}
    \EndIf
    \State $\text{TxTime}[\mathbf{w}[k].\text{user}]{=}\text{NewTime}[\mathbf{w}[k].\text{user}]$
    \EndFor
    \State \textbf{return} $\omega_{\tau}$, $\rho_{\tau}$
\end{algorithmic}
\end{algorithm}

\vspace{4pt}$\bullet$ The algorithm first calculates the $\mathbf{r}$-weighted index values $W_i^{\mathbf{r}}(\pi_i)$ by scaling $W_i^{\mathbf{1}}(\pi_i)$ by $r_i$, and stores the value and the corresponding user in vector $\mathbf{I}$ (line 7-15).

$\bullet$ The algorithm then sorts all the $\mathbf{r}$-weighted indices of each belief state of all users to a $(2\tau{+}1)N$-dimensional vector $\mathbf{w}$ in increasing order (line 16).

$\bullet$ The algorithm then calculates $\omega_{\tau}$ and $\rho_{\tau}$ based on the monotonicity property in Lemma~\ref{lemma:alpha_tau}(ii). Hence, fixing the randomization factor $\rho{=}1$, it increases the threshold $\omega$ by going through the indices in $\mathbf{w}$ and calculates the long-term average number of transmission when threshold $\omega$ equals to that index. For each element of $\mathbf{w}$, it first calculates the long-term expected fraction of time $\text{NewTime}[\mathbf{w}[k].\text{user}]$ transmitting to the corresponding user $\mathbf{w}[k].\text{user}$ in line 18, and hence the decreased amount, denoted by TimeDiff, as compared with previous value $\text{TxTime}[\mathbf{w}[k].\text{user}]$ in line 19. Note that, in each iteration, only the user corresponding to $\mathbf{w}[k]$ will have an updated expected fraction of transmission time. The total expected number of transmission, denoted by TotalTime, is then updated by decreasing the same amount (line 20). The threshold $\omega$ keeps increasing until the total expected number of transmission is below $M$ (line 21). Noting that  ${\alpha}^{\tau}_i(\omega, 1)$ decreases with $\omega$, we then set $i_{\tau}=\mathbf{w}[k-1].\text{user}$ and $\omega_{\tau}=\mathbf{w}[k-1].\text{value}$ (line 21-22). Then we calculate the expected transmission time to the user that corresponds to $\mathbf{w}[k-1]$ (line 23) and select the randomization factor $\rho_{\tau}$ so that the constraint (\ref{eq:tau_constr}) is satisfied (line 24), where the function $\beta_i: (\omega, \alpha)\rightarrow \rho$ calculates the randomization factor $\rho$ required to achieve the long-term expected fraction of time $\alpha$ transmitting to user $i$ at threshold $\omega$, and is derived from lemma~\ref{lemma:alpha_tau}(i) as,
\begin{align}
&\beta_i(\omega,\alpha)\nonumber\\
={\hspace{-5pt}}&\begin{cases}
\frac{(1-\alpha)(1-p^i_{11}+b^i_{0,h+1})-\alpha h(1-p^i_{11})}{(1-\alpha) (b^i_{0,h+1}-b^i_{0,h})-\alpha(1\hspace{-1pt}{-}\hspace{-1pt}p^i_{11})} &\text{if $\omega{=}W^{\mathbf{r}}_i(b^i_{0,h})$, $h{<}\tau$;}\\
\frac{(1-\alpha)(1-p^i_{11}+b^i_{s})-\alpha \tau(1-p^i_{11})}{(1-\alpha) (b^i_{s}-b^i_{0,\tau})-\alpha(1\hspace{-1pt}{-}\hspace{-1pt}p^i_{11})} &\text{if $\omega{=}W^{\mathbf{r}}_i(b^i_{0,\tau})$;} \\
\frac{\alpha(1-p^i_{11})}{(1-\alpha\tau)(1-p^i_{11})+(1-\alpha) b^i_{s}} &\text{if $\omega{=}W^{\mathbf{r}}_i(b^i_s)$;}\\
0 &\text{if $\omega{>}W^{\mathbf{r}}_i(b^i_s)$.}
\end{cases}\nonumber
\end{align}

\vspace{-5pt}\subsection{Performance of policy over untruncated state space with approximate parameters $\omega_{\tau},\rho_{\tau}$}
\label{sec:thr_appro_alg}

We next examine, over the \emph{original untruncated model}, the policy that uses the approximated parameters $i_{\tau}$, $\omega_{\tau}$ and $\rho_{\tau}$. We denote such policy as $\phi_{\tau}(\mathbf{r}, M)$ and present it next.

\begin{algorithm}
  \caption{\hspace{-4pt}$\phi_{\tau}(\mathbf{r}, M)$\textbf{:} $\mathbf{r}$-weighted Index Policy}\label{alg:thres}
  \begin{algorithmic}[1]
  \State \textbf{Initialization phase:} The parameters $i_{\tau}$, $\omega_{\tau}$ and $\rho_{\tau}$ are calculated by algorithm $G^{\tau}(\mathbf{r},M)$.
  \State \textbf{At slot $\bm t$:} user $i$ is scheduled if the $\mathbf{r}$-weighted index value $W^{\mathbf{r}}_i(\pi_i[t])> \omega_{\tau}$, or if $W^{\mathbf{r}}_i(\pi_i[t]){=}\omega_{\tau}$ with $i{>}i_{\tau}$. User $i$ stays passive if \hspace{3pt}$W^{\mathbf{r}}_i(\pi_i[t])<\omega_{\tau}$, or if $W^{\mathbf{r}}_i(\pi_i[t])\hspace{1pt}{=}\hspace{1pt}\omega_{\tau}$ with $i{<}i_{\tau}$. If $W^{\mathbf{r}}_i(\pi_i[t]){=}\omega_{\tau}$  with $i{=}i_{\tau}$, user $i$ is
scheduled with probability $\rho_{\tau}$.
\end{algorithmic}
\end{algorithm}

\vspace{3pt}\noindent\textbf{Remark:} The computational complexity of the initialization phase of algorithm $\phi_{\tau}(\mathbf{r}, M)$ is dominated by sorting the index values in Algorithm $G^{\tau}(\mathbf{r},M)$ (line 16), which has complexity $O\big((2\tau+1)N \cdot \log \big((2\tau+1)N\big)\big)$.
After initialization, the $\mathbf{r}$-weighted Index Policy $\phi_{\tau}(\mathbf{r}, M)$ takes a very simple threshold-type form with per-slot computational complexity $O(N)$.

We let $V^*(\mathbf{r},M)$ be the weighted sum-throughput under the optimal policy $\phi^*(\mathbf{r}, M)$ defined in lemma~\ref{lemma:thres_index}, and let $V_{\tau}(\mathbf{r},M)$ be that under the afore-mentioned policy $\phi_{\tau}(\mathbf{r}, M)$, i.e.,
\begin{align}
V^*(\mathbf{r},M){=}\liminf_{T\rightarrow \infty} \frac{1}{T} \mathbb{E}\Big[\sum_{t=0}^{T-1} \sum_{i=1}^N r_i {\cdot} \pi_i[t] {\cdot} a_i^{\phi^*(\mathbf{r}, M)}[t] \Big].\label{eq:thr_nontrun}\\
V_{\tau}(\mathbf{r},M){=}\liminf_{T\rightarrow \infty} \frac{1}{T} \mathbb{E}\Big[\sum_{t=0}^{T-1} \sum_{i=1}^N r_i {\cdot} \pi_i[t] {\cdot} a_i^{\phi_{\tau}(\mathbf{r}, M)}[t] \Big].\label{eq:thr_trun}
\end{align}

Since we also require the long-term average number of transmissions of the policy $\phi_{\tau}(\mathbf{r}, M)$ to satisfy the constraint~(\ref{eq:constraint}), we denote $Z_{\tau}(\mathbf{r},M)$ as the time-average expected number of transmissions under this policy, i.e.,
\begin{align}
Z_{\tau}(\mathbf{r},M)=\limsup_{T\rightarrow \infty} \frac{1}{T} \mathbb{E}\Big[ \sum_{t=0}^{T-1}\sum_{i=1}^N a_i^{\phi_{\tau}(\mathbf{r},M)}[t]\Big]. \nonumber
\end{align}

Recall that $\tau_0$ is defined in Lemma~\ref{lemma:alpha_tau}. The next lemma shows that the policy $\phi_{\tau}(\mathbf{r}, M)$ asymptotically achieves the maximum weighted sum-throughput of (\ref{eq:obj_rel})(\ref{eq:cons_rel}) as the truncation size increases, while abiding the long-term average number of transmissions constrain~(\ref{eq:constraint}). The proof is given in Appendix~\ref{appen:f_tau}.

\vspace{5pt}\begin{lemma}\label{lemma:eps_bound_tau}
\vspace{-9pt}\noindent For $\tau\geq\tau_0$, we have

\noindent(i) The weighted sum-throughput performance difference between the policies $\phi^*(\mathbf{r}, M)$ and $\phi_{\tau}(\mathbf{r}, M)$ is bounded by
\begin{align}
|V^*(\mathbf{r},M)-V_{\tau}(\mathbf{r},M)| \leq f(\tau) \sum_{i=1}^N r_i,
\end{align}

\vspace{-6pt}\noindent where $f(\tau){=}\sum_{i=1}^N f_i(\tau)$, which satisfies $f(\tau){\rightarrow} 0$ as $\tau{\rightarrow}\infty$ with
\begin{align}
\label{eq:act_time0}
f_i(\tau)=\frac{\rho(b^i_{0,\tau}-b^i_{0,\tau+1})+1-p^i_{11}+b^i_{0,\tau{+}1}}{\rho b^i_{0,\tau}{+}(1{-}\rho)b^i_{0,\tau\hspace{-1pt}{+}\hspace{-1pt}1}{+}(1\hspace{-1pt}{-}\hspace{-1pt}p^i_{11})(\tau{+}1{-}\rho)}.
\end{align}

\noindent(ii) The long-term average number of transmissions under policy $\phi_{\tau}(\mathbf{r}, M)$ satisfies the constraint~(\ref{eq:constraint}), i.e., $Z_{\tau}(\mathbf{r},M)\leq M$.
\end{lemma}

\noindent\textbf{Remark:} Note that the truncation size $\tau$ needs to be sufficiently large (i.e., $\tau\geq\tau_0$) to prove the Lemma. This is because sufficiently large truncation size can provide enough level of approximation that facilitates analytical characterization. Specifically, in the proof, $\tau_0$ is used in Lemma~\ref{lemma:sublinear}.

\vspace{8pt}\section{QUEUE-BASED INDEX POLICY OVER TIME FRAMES}
\label{sec:QWI_policy}

Note that the Index Policy in the last section, as well as the associated Whittle's index value, is for the system with infinitely backlogged queues and the corresponding weighted sum-throughput maximization problem~(\ref{eq:obj_rel})-(\ref{eq:cons_rel}). In this section, we consider scheduler design under random arrival of data packets and the associated queue evolution in the time-correlated downlink. The objective here is to not only obtain maximum weighted sum-throughput, but also maintain queue stability. In the presence of queue evolution, the problem get much more complicated. Note that, in the weighted sum-throughput maximization problem, the reward of scheduling a user is captured by the Whittle's index value. Under the additional consideration of queue stability, the queue lengths need to be jointly taken into account for scheduling, i.e., a user is scheduled for transmission not only because it has a high index value, but may also because it has a large queue length.

Next, we propose a throughput-optimal scheduling policy based on scaling the Whittle's index by the queue length. The policy is implemented over separate time-frames and has low-complexity.

We divide the time slots $\{0,1,2,\cdots \}$ into separate \emph{time frames} of length $T$, i.e., the $k$-th frame,  $k\in \{0,1,2,\cdots\}$, includes time slots $kT, {\cdots}, (k+1)T{-}1$. The scheduling decisions in the $k$-th frame are made based on the queue length information $\mathbf{q}[kT]$ at the beginning of that frame. During the $k$-th frame, the policy $\phi_{\tau}(\mathbf{q}[kT],M)$, developed in the last section, is implemented. Formally, the $T$-frame queue-based index policy, denoted by $\text{Q-Index}_{\tau}\text{(T,M)}$, is introduced next.

\vspace{-5pt}\begin{algorithm}[H]
  \caption{\hspace{-5pt}$\text{Q-Index}_{\tau}\hspace{-1pt}\text{(T,M)}$\textbf{:} \hspace{-2pt}$T$-Frame \hspace{-1pt}Queue-based \hspace{-1pt}Index \hspace{-1pt}Policy}\label{alg:thres}
  \begin{algorithmic}[1]
  \State The time slots are divided into frames of length $T$. Slot $t$ is in the $k^{th}$ frame if $kT\leq t < (k+1)T$, $k\in\{0, 1, \cdots\}$.
  \State \textbf{At the beginning of the $\bm k^{th}$ frame:} At the beginning of slot $kT$, implement the algorithm $G^{\tau}(\mathbf{q}[kT],M)$ that outputs $\omega_{\tau}$ and $\rho_{\tau}$.
  \State \textbf{In each slot $\bm t$ of the $\bm k^{th}$ frame:}
  \Statex$\bullet$\textbf{User scheduling:} user $i$ is scheduled if the $\mathbf{q}[kT]$-weighted index value $W^{\mathbf{q}[kT]}_i(\pi_i[t]){>}\omega_{\tau}$, or if $W^{\mathbf{q}[kT]}_i(\pi_i[t]){=}\omega_{\tau}$ with $i{>}i_{\tau}$. User $i$ stays passive if $W^{\mathbf{q}[kT]}_i(\pi_i[t]){<}\omega_{\tau}$, or if $W^{\mathbf{q}[kT]}_i(\pi_i[t]){=}\omega_{\tau}$ with $i{<}i_{\tau}$.
If $W^{\mathbf{q}[kT]}_i(\pi_i[t])=\omega_{\tau}$ with $i{=}i_{\tau}$, user $i$ is
scheduled with probability $\rho_{\tau}$. If a user with empty queue is scheduled, then a dummy packet is transmitted to the user.
    \Statex$\bullet$\textbf{ARQ feedback:} At the end of each slot, the scheduled users send ARQ feedback to the BS. The belief values are updated according to the feedback at the scheduler.
\end{algorithmic}
\end{algorithm}

\noindent \textbf{Remarks}: We next describe the intuition behind designing the above algorithm.

\noindent(1) Note that, for queue stability, instead of using queue length information in every slot, it is sufficient only to consider the sampled queue length information at the periodic slots, i.e., $\mathbf{q}[kT], k=0,1,\cdots$. The queue is stable if and only if the periodically sampled queue length evolution process is stable.

\noindent(2) Within each frame, we wish to maximize the weighted sum-throughput, where each user's throughput is weighted by its queue length sample value at the beginning of the time frame. Hence, in step 2-3, we implement the Index policy $\phi^{\tau}(\mathbf{q}[kT],M)$ developed in the previous section. The rationale is because, first, we would like to schedule the users to achieve the higher throughput promised by the Index policy that exploits the temporal correlated channels. Moreover, for system stability, we would like to choose users with large queue-lengths. Hence, by considering the queue weighted throughput and using the Index policy $\phi^{\tau}(\mathbf{q}[kT],M)$ in frame $T$, an overloaded queue can get served with potentially higher rate. As a direct result, a user $i$'s index is scaled by its queue length $q[kT]$.

\noindent(3) An intuitive explanation of the multiplication of index and queue length is as follows. We schedule a user not only because of its longer queues, but also when its underlying `channel quality' is favorable (in terms of both exploitation and exploration values). Consider the example where a user's channel is strongly correlated and is observed `0' state in the previous slot. Hence it is highly likely to stay in `0' state for a while. Hence scheduling it can result in wasted system resource since packets are unlikely to be successfully delivered. Correspondingly, this `quality' of a channel is reflected in the close-to-zero Whittle's index value.  The multiplication of queue length and the Whittle's index value is able to capture both the queue length and the channel's `quality' for scheduling. Summation of the index and queue length, on the other hand, fails capture both of these properties.

\noindent(4) Dividing the time slots into different frames brings us advantages in the realm of large frame length (i.e., $T$). Since we implement the Index policy within each finite-horizon frame, if the frame length is small, we lose from exploiting the channel correlation because the Index policy is optimal only in the infinite horizon. As  the frame length scales, the (per-slot) loss of exploiting the channel correlation diminishes.

\noindent(5) Note that a dummy packet is transmitted to a scheduled user with empty queue. The dummy packet is known to the users and contains no new information
and hence does not bring throughput gains if it is transmitted. However, the scheduler will still receive channel state update from the corresponding scheduled users. This mechanism is useful to establish our results.

\vspace{3pt}The next proposition and corollary establish throughput-optimality of the queue-based index policy over time frames, where, recall that, $f(\tau)$ is given in Lemma~\ref{lemma:eps_bound_tau}. The proof is given in Appendix~\ref{appen:thr_opt}.

\begin{proposition}\label{prop:tau_bound}
If $\tau{\geq} \tau_0$, then there exist $T_0$ and function $g(\tau){=}3f(\tau)$ such that the following holds whenever $T{>}T_0$: If the arrival rate $\mathbf{\bm\lambda}$ satisfies $\mathbf{\bm\lambda}+g(\tau) \mathbf{1}{\in}\bm \Gamma$ and the $T$-frame queue-based index policy $\text{Q-Index}_{\tau}(T,M{-}g(\tau)/2)$ is implemented, then all queues are stable and constraint ($\ref{eq:constraint}$) on the average number of transmissions is satisfied.
The function $g(\tau)$ satisfies $\lim_{\tau \rightarrow \infty}g(\tau)=0$.
\end{proposition}

\vspace{6pt}

\begin{corollary}\label{cor:GammaLambda}
The achievable rate region $\bm\Gamma$, expressed in (\ref{eq:rate_reg}), is equal to the stability region $\bm\Lambda$.
\end{corollary}

\noindent \textbf{Proof:} Recall that the achievable rate region $\bm \Gamma$ corresponds to the expected service rate vectors that can be achieved in the system with infinitely backlogged queues, by any policy in $\Phi$. Now consider all the arrival rates within the interior of the stability region $\bm\Lambda$. For each arrival vector $\mathbf{\bm\lambda} \in \bm \Lambda$, there exists a certain policy in $\Phi$ that stabilizes it, i.e., provides a service rate not below $\mathbf{\bm\lambda}$. Therefore, the achievable rate region $\Gamma$ provides an upper bound on the stability region $\bm \Lambda$. Since the previous proposition states that the queue-based index policy stabilizes arrival rates arbitrarily close to the boundary of the achievable rate region $\bm \Gamma$, the achievable rate region $\bm \Gamma$ and the stability region $\bm \Lambda$ share the same interior. Because both regions $\bm \Gamma$ and $\bm \Lambda$ are defined over closure of sets, we have $\bm \Gamma=\bm \Lambda$. \hfill $\blacksquare$
\vspace{4pt}

Proposition~\ref{prop:tau_bound} and Corollary~\ref{cor:GammaLambda} together establish the throughput optimality of the proposed policy. With sufficiently large $\tau$ and $T$, the proposed policy  $\text{Q-Index}_{\tau}(T,M{-}g(\tau)/2)$ can support arrival rate $\bm\lambda$ within arbitrary $\epsilon$ interior of the stability region, i.e.,  $\bm\lambda+\epsilon\mathbf{1} \in\mathbf{\Lambda}$ and satisfy constraint~(\ref{eq:constraint}).

\noindent\textbf{Remarks:}

\noindent(1) Note that, in Proposition~\ref{prop:tau_bound}, the parameter $M$ in the queue-based index policy is scaled down by $g(\tau)/2$. This mechanism is needed to guarantee the constraint on the long-term average number of transmission. The details are given in the proof.

\noindent(2) In the queue-based index policy, a user is scheduled based on its $\mathbf{q}[kT]$-weighted Whittle's index value. The Whittle's index value is necessary for the results because it measures the importance of a wireless channel for scheduling, considering jointly the instantaneous throughput and future throughput (e.g., see \cite{Zhao_index}\cite{Whittle}, Lemma~\ref{lemma:thres_index}). It is interesting to note that a simple multiplication of queue length and Whittle's index value captures the importance of scheduling a user under two sophisticated system features -- the queue evolution and the fundamental exploration-exploitation tradeoff.

\noindent(3) Calculation of $\mathbf{q}[kT]$-weighted index value is very simple, which only requires scaling the \emph{pre-calculated} Whittle's index value. Under the queue-based index policy, in each frame, implementation of $G^{\tau}(\mathbf{q}[kT],M{-}g(\tau))$ in step $2$ of policy $\text{Q-Index}_{\tau}(T,M{-}g(\tau))$ has computational complexity $O((2\tau+1)N \log (2\tau+1)N)$, while implementing step $3$ of policy $\text{Q-Index}_{\tau}(T,M{-}g(\tau))$ over the frame has complexity $O(T N)$ (see the remark in Section \ref{sec:thr_appro_alg}). Accordingly, the \emph{per-frame} complexity is $O((2\tau+1)N \log (2\tau+1)N+TN)$. Therefore, as the frame length $T$ scales up, the \emph{per-slot} complexity decreases toward $O\big(N\big)$.

\noindent(4) The scheduling decisions are made by comparing each user's own index value to a threshold, independently from other users. Hence our policy is also applicable for \emph{distributed implementation} in uplink scenarios.

\begin{figure}
\centering
\includegraphics[width=3.6in]{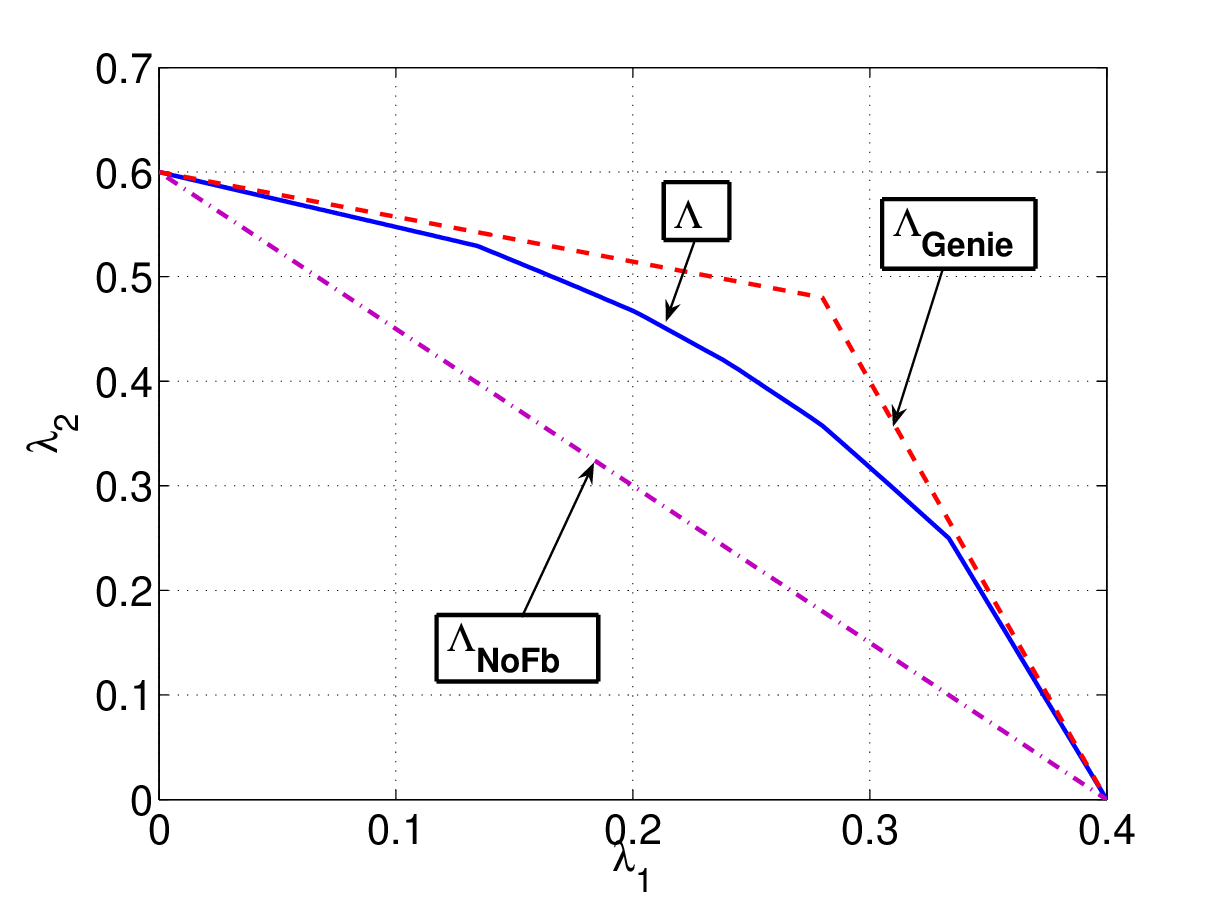}
\caption{Comparison of stability regions. Parameters used: $p^1_{11}=0.7$, $p^1_{01}=0.2$; $p^2_{11}=0.8$, $p^2_{01}=0.3$}
\label{fig:reg_cmpr}\vspace{-8pt}
\end{figure}

\section{Numerical Results}
\label{sec:num}

\subsection{Illustration of Stability Region}
\label{sec:num:region}

In Fig.~\ref{fig:reg_cmpr}, we compute the stability region $\bm \Lambda$ and
compare it with other regions of interest. We consider the scenario with two users and with the scheduling constraint on the long-term average number
of scheduled transmissions $M=1$. The Markov transition statistics are selected as $(p^1_{11},p^1_{01})=(0.7,0.2)$, $(p^2_{11},p^2_{01})=(0.8,0.3)$. For comparison, in the same system, we consider another scenario where the scheduler throws away the ARQ feedback from the scheduled user. We denote the corresponding stability region by $\bm \Lambda_{NoFb}$, expressed as $\bm\Lambda_{NoFb}=\{\bm \lambda: \lambda_1/b^1_s+\lambda_2/b^2_s\leq 1\}$ \cite{Neely_blind}. As can be observed in the figure, by exploiting the channel memory from ARQ feedback, our policy achieves significant throughput gain (as high as $30\%$) over the policy that ignores the channel memory. We also compare the stability region $\bm \Lambda$ with that of a `genie-aided' system, denoted by $\bm \Lambda_{Genie}$. In the `genie-aided' system, the same scheduling constraint (\ref{eq:constraint}) is imposed, while a genie reveals channel states of \emph{all users} in the current slot to the scheduler at the end of the slot. The region $\bm\Lambda_{Genie}$ is expressed as
\begin{align}
\bm\Lambda_{Genie}=b^1_s b^2_s \bm\lambda_{00}+(1-b^1_s) b^2_s \bm\lambda_{01}+b^1_s (1-b^2_s)\bm\lambda_{10}\nonumber\\
\hspace{1.2in}+(1-b^1_s)(1-b^2_s)\bm\lambda_{11}\nonumber
\end{align}
with $\bm\lambda_{ij}\in \bm\Lambda_{ij}$ where $\bm\Lambda_{ij} = \mathcal{CH} \{(p^1_{i1}, 0), (0, p^2_{j1})\}$, $i,j=0,1$  with $\mathcal{CH}\{\cdot\}$ denoting the convex hull of the set \cite{KrishnaModiano}.
Because the genie facilitates more informed decisions at the scheduler, the resultant stability region $\Lambda_{Genie}$ provides an outer bound on region $\Lambda$, as demonstrated in Fig.~\ref{fig:reg_cmpr}.

\subsection{Delay Performance Analysis}

In this section, we numerically evaluate the delay performances of the proposed policy. We consider a two users system with the long-term average number of transmission constraint $M=1$, i.e., one user can be scheduled on average. The channel states of both users evolve as the `ON/OFF' Markov chain with transition statistics $(p^1_{11},p^1_{01}){=}(0.7,0.2)$, $(p^2_{11},p^2_{01}){=}(0.8,0.3)$, i.e., which can be typical situations where both users have moderate degree of correlation across time.

\begin{figure}
\centering
\includegraphics[width=3.6in]{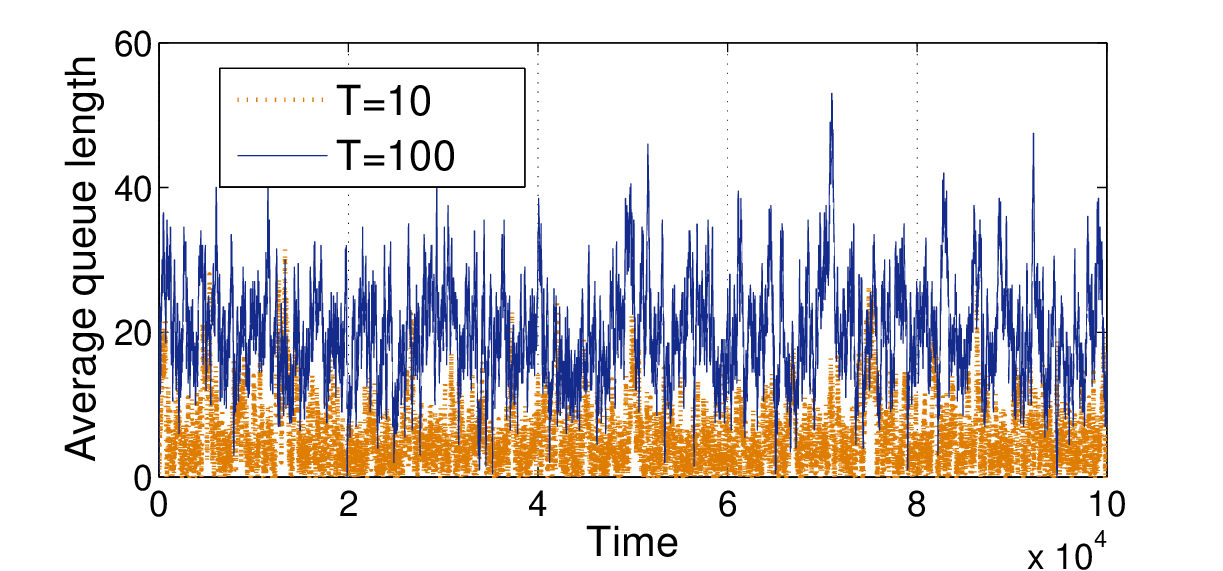}
\caption{Sample paths of queue evolution.}
\label{fig:Delay_sample}\vspace{-8pt}
\end{figure}

Over this system, we implement the proposed $T$-frame queue-based index policy $\text{Q-Index}_{\tau}(T,M{-}g(\tau)/2)$, defined in section~\ref{sec:QWI_policy} with $\tau{=}20$. We first consider fixed arrival rates $\lambda_1{=}\lambda_2{=}0.25$ and implement the policies $\text{Q-Index}_{\tau}(T,M{-}g(\tau)/2)$ with frame lengths $T{=}10$ and $T{=}100$, respectively. The sample paths of the average queue length, i.e., $\big(Q_1[t]+Q_2[t]\big)/2$, are plotted in Fig.~\ref{fig:Delay_sample}. It can be observed that, while the queues in both scenarios are stable, the variation of the queue evolution is notably higher when the frame size changes from $10$ to $100$. This is because, as the frame size increases, the frame-based algorithm obtains less frequent updates of the queue sizes. Therefore, within a frame, the algorithm can continue to serve a user even if its current queue length becomes small while neglecting the other user that has accumulated a large queue size, leading to a higher degree of queue length variation as well as average queue size. Correspondingly, higher delay and delay variation are expected as the frame size increases. For example, suppose the initial queue length of user $1$ is empty, while the initial queue length for user $2$ is nonempty. Then user $1$ in the first frame will not be scheduled. Now after the first frame, the expected queue length of user $1$ will be significantly larger for the case when $T=100$ compared with the case when $T=10$. Hence, at the second frame, the scheduler dedicates most of the resources to user $1$. As a result, the expected queue length of the user $1$ will go down after second frame, and the expected queue length of user $2$ will grow. Both the expected change of queue lengths of user $1$ and $2$ will be much more significant when $T=100$ compared with when $T=10$. The process repeats in time and results in a higher degree of queue length variation when $T=100$ as compared to $T=10$.

\begin{figure}
\centering
\includegraphics[width=3.5in]{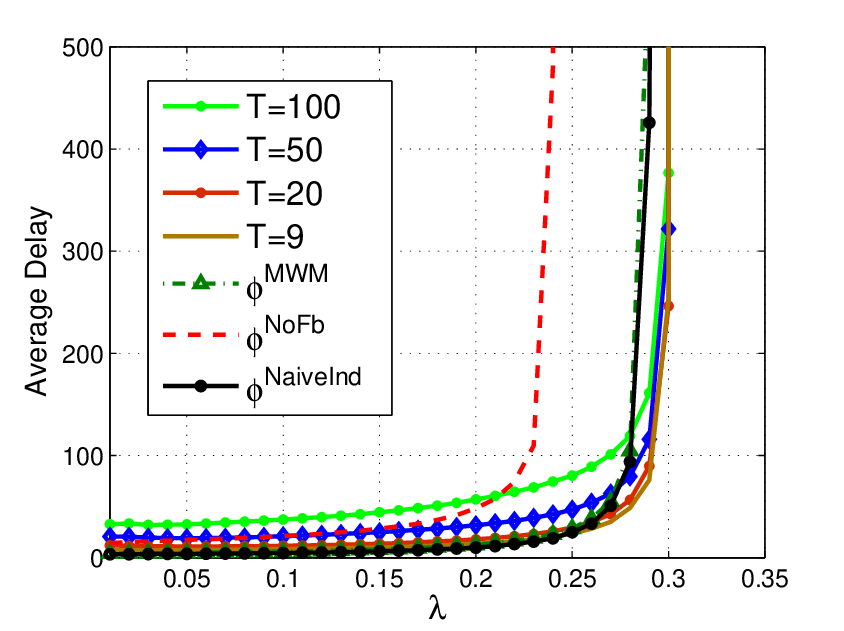}
\caption{Delay performance comparison when $N=2$. }
\label{fig:Delay}
\end{figure}

We next implement the aforementioned policy $\text{Q-Index}_{\tau}(T,M{-}g(\tau)/2)$ and evaluate the average queueing delay experienced by users as the arrival rates scale toward the boundary of the stability region, with varying frame length $T$. For the two user system previously discussed, Fig.~\ref{fig:Delay} examines the average queueing delay when the arrival rate vector $(\lambda_1,\lambda_2)$ increases with $\lambda_1{=}\lambda_2{=}\lambda$. As can be observed in the figure, as the arrival rates grow toward the boundary of stability region, the queue length quickly blows up, resulting in steep increase of delay. The steep increase is because, as the arrival rates grow toward the boundary of stability region, the queue lengthes quickly blow up because they are becoming unstable, resulting in steep increase of average delay. Fig.~\ref{fig:Delay} also show that, as the frame length grows, the average delay in the downlink network increases. This is, again, a consequence of infrequent update of queue length information at the scheduler.

Another interesting observation can be observed from Fig.~\ref{fig:Delay}. When we implement the proposed policy $\text{Q-Index}_{\tau}(T,M{-}g(\tau)/2)$ with the frame lengths $T$ growing from $9$ to $100$, the system delay curves for different values of $T$ start to build up significantly at \emph{around the same value} (i.e., around $0.29$ which is on the boundary of the stability region). Note that we needed the frame size to be large enough to prove Proposition~\ref{prop:tau_bound}. However, in practice, the frame size $T$ may not need to be as large to guarantee queue stability. This numerical result, along with many other numerical evaluations we have conducted, indicates that the queues are stable under only moderate value of frame size in the proposed queue-based index policy.

Fig.~\ref{fig:Delay} also plots the delay performance of a policy $\phi^{NoFb}$ that ignores the channel memory, i.e., not using the channel state feedback. In each slot of this policy, a user $i$ with the largest multiplication of steady state transmission rate (i.e., $b^i_{s}$) and queue length $q_i[t]$ is scheduled. The delay performance of maximum weight matching policy $\phi^{MWM}$ is also plotted, where, in each slot $t$, a user $i$ with the largest multiplication of belief value $\pi^i[t]$ and queue length $q_i[t]$ is scheduled.
Fig.~\ref{fig:Delay} further plots the delay performance of a naive policy $\phi^{NaiveInd}$ where a user $i$ with the largest multiplication of index value $W^1_i(\pi^i[t])$ and queue length $q_i[t]$ is scheduled. For all of these policies, the values of arrival rate $\lambda$ where the queueing delay increases steeply are at a smaller value than our proposed policy, implying the sub-optimality of these policies. This is partly because these policies only schedule strictly $M$ users per slot, but our work is in the domain of a relaxed constraint of average number of scheduled users. The sub-optimality of policy $\phi^{MWM}$ is also because it only exploits the channel condition in the instantaneous slot, i.e., $\pi_i[t]$, but it does not consider exploring outdated channels. It is interesting to note that policy $\phi^{MWM}$ and $\phi^{NaiveInd}$ performs better than the policy that ignores channel state feedback, as the value of $\lambda$ where steep increase of queueing delay occurs is much larger as compared to $\phi^{NoFb}$. This observation illustrates the throughput gains that can be achieved by using the channel state feedback.
\section{Conclusion}

In this work, we have studied downlink scheduling problem over Markovian evolving ON/OFF fading channels and imperfect instantaneous channel state information. The scheduling decisions are made based on the single-bit ARQ-type feedback and the channel memory inherent in the Markovian channels. We propose a throughput-optimal policy that operates over time frames. In the proposed policy, the importance of scheduling a user is measured by a simple multiplication of the queue length and Whittle's index value. Because of this property, the proposed policy has low-complexity per frame in the network size and the truncation level of the belief state space. Most notably, our policy does not suffer from the curse of dimensionality that is observed in earlier works in this context. Numerical evaluations show that significant throughput performance gains can be achieved by exploiting the channel memory, via the frame-based low-complexity queue-based index policy with moderate frame size. Future directions include considering larger state space model, and considering feedback mechanisms that collects CSI from unscheduled users, as well as more stringent instantaneous scheduling constraints. Another open direction is to consider adaptive power allocation with hybrid ARQ protocols (e.g., \cite{Green_HARQ}), where the index value not only implies the attractiveness of scheduling a user, but also guides the power allocation across time.

\appendices

\vspace{-0.1in}\section{Proof of Lemma~\ref{lemma:thres_index}}
\label{appen:threshold}

The proof of the lemma is an extension of the proof of Proposition~1 in \cite{Wenzhuo_infocom12}. Consider the problem $\Psi(\mathbf{r},M)$ with weight vector $\mathbf{r}$. The constraint (\ref{eq:constraint}) can be written in an equivalent form that requires at least $N-M$ channels to be \emph{passive} on average, i.e.,
\begin{align}
\label{eq:equiv_ineq}
\liminf_{T\rightarrow \infty} \frac{1}{T} \mathbb{E}\Big[\hspace{-1pt}\sum_{t=0}^{T-1} \sum_{i=1}^{N} (1{-}a_i^{\phi}[t]) \Big ] \geq N-M.
\end{align}

Associating a Lagrange multiplier $\omega$ to the constraint (\ref{eq:equiv_ineq}), we have the following Lagrangian function $L(\phi,\omega)$ for problem $\Psi(\mathbf{r},M)$,
\begin{align}
\label{eq:Lagrange}
L(&\phi, \omega){=}\liminf_{T\rightarrow \infty} \frac{1}{T} \mathbb{E}\Big[\sum_{t=0}^{T-1} \sum_{i=1}^N r_i {\cdot} \pi_i[t] {\cdot} a_i^{\phi}[t]\Big] \nonumber \\
&{+}\omega {\cdot} \liminf_{T\rightarrow \infty} \frac{1}{T} \mathbb{E}\Big[\sum_{t=0}^{T-1} \sum_{i=1}^N  (1{-}a^{\phi}_i[t])\Big]{-} \omega {\cdot} (N{-}M) .
\end{align}

The dual function $D(\omega)$ is defined as $D(\omega)=\max_{\phi\in \Phi} L(\phi,\omega)$. Following the lines of proof in \cite{Wenzhuo_infocom12} we have
\begin{align}
D(\omega)=\sum_{i=1}^N U_i^{r_i}(\omega)+\omega (N-M). \nonumber
\end{align}
in which $U_i^{r_i}(\omega)$ is a $\omega$-subsidy problem under weight $r_i$,
\begin{align}
U_i^{r_i}(\omega)&= \max_{\phi \in \Phi_i} \limsup_{T\rightarrow \infty} \frac{1}{T} \mathbb{E}\Big[\sum_{t=0}^{T-1}\big[r_i{\cdot}\pi_i[t] {\cdot} a^{\phi}_i[t] \nonumber \\
&\hspace{1.5in} +\omega \cdot (1{-}a^{\phi}_i[t]) \big]\Big],\label{eq:v_sub}
\end{align}
where $\Phi_i$ denotes the set of scheduling policies that activate and idle the user $i$ according to the observed channel history. In the above problem (\ref{eq:v_sub}), for each channel $i$ at belief state $\pi_i$, it will receive a reward $r_i \pi_i$ when it activates, otherwise it will receive a subsidy $\omega$ for passivity. We let $\mathcal{I}^{r_i}_i(\omega) \subseteq \mathcal{B}_i$ be the set of belief states for which it is optimal to stay idle.

Under the unit weight $r_i=1$, it was shown in
\cite{Zhao_index} that the problem is Whittle indexable, i.e., $\mathcal{I}^{1}_i(\omega)$ monotonically increases from $\emptyset$ to $\mathcal{B}_i$ as $\omega$ increase from $0$ to $\infty$ for each user $i$. The Whittle's index value $W_i^{\mathbf{1}}(\pi)$ is defined as the infimum subsidy value for which the belief state $\pi$ is at the boundary of $\mathcal{I}^{1}_i(\omega)$, i.e.,
\begin{align}
W_i^{\mathbf{1}}(\pi)=\inf \{ \omega: \pi \in \mathcal{I}^{1}_i(\omega) \}. \nonumber
\end{align}

It follows from \cite{Wenzhuo_infocom12} that, for the $\omega$-subsidy problem under unit weight $r_i=1$, the optimal policy is to activate the user at time slot $t$ if $W_i^{\mathbf{r}}(\pi)>\omega$, and to stay idle if $W_i^{\mathbf{r}}(\pi)<\omega$, with tie breaking arbitrarily if $W_i^{\mathbf{r}}(\pi)=\omega$.

We next extend the optimal algorithm for the $\omega$-subsidy problem under unit weight to the general case with arbitrary non-negative weight $r_i$. An equivalent form of $U_i^{r_i}(\omega)$ is as follows,
\begin{align}
&U_i^{r_i}(\omega) \nonumber\\
\hspace{-3pt}{=}&r_i  \max_{\phi \in \Phi_i} \limsup_{T\rightarrow \infty} \frac{1}{T} \mathbb{E}\Big[\sum_{t=0}^{T-1}\big[\pi_i[t]  a^{\phi}_i[t] {+} \frac{\omega}{r_i}  (1{-}a^{\phi}_i[t]) \big]\Big].\label{eq:v_Rsub}
\end{align}

Therefore, the optimal solution for the $\omega$-subsidy problem (\ref{eq:v_sub}) with weight $r_i$ takes the same form as the optimal solution for the $\omega/r_i$-subsidy problem with weight $1$. Accordingly, the optimal solution takes the following form: a user $i$ is scheduled at slot $t$ if $W_i^{\mathbf{r}}(\pi_i[t])>\omega/r_i$, and stay idle if $W_i^{\mathbf{r}}(\pi)<\omega/r_i$, with tie breaking arbitrarily if $W_i^{\mathbf{1}}(\pi)=\omega/r_i$.

We define the $\mathbf{r}$-weighted index value as $W_i^{\mathbf{r}}(\pi)=r_i \cdot W_i^{\mathbf{1}}(\pi), \pi \in \mathcal{B}_i, i \in \{1,\cdots, N\}$. The optimal policy for the reward maximization problem in (\ref{eq:v_Rsub}) is then to activate the user $i$ if $W_i^{\mathbf{r}}(\pi)>\omega$, and to stay idle if $W_i^{\mathbf{r}}(\pi)<\omega$, with tie breaking arbitrarily if $W_i^{\mathbf{r}}(\pi)=\omega$. Because of this threshold-based policy and arbitrary tie-breaking at the threshold, the dual function value $D(\omega)$ can be achieved by the following threshold-based policy implemented over the $\mathbf{r}$-weighted index values $W_i^{\mathbf{r}}(\pi)$: User $i$ is scheduled if $W^{\mathbf{r}}_i(\pi_i){>}\omega$, or if $W^{\mathbf{r}}_i(\pi_i){=}\omega$ with $i{>}j$. User $i$ stays idle if \hspace{3pt}$W^{\mathbf{r}}_i(\pi_i){<}\omega$, or if $W^{\mathbf{r}}_i(\pi_i){=}\omega$ with $i{<}j$.
If $W^{\mathbf{r}}_i(\pi_i){=}\omega$ with $i{=}j$, user $i$ is scheduled with probability $\rho$.

Following the similar proof techniques of Lemma 11 in \cite{Wenzhuo_infocom12}, by appropriately choosing the aforementioned parameters $(j,\omega,\rho)$ to be $(i^*,\omega^*,\rho^*)$ such that the constraint~(\ref{eq:constraint}) on the average number of transmissions is strictly satisfied with equality, the corresponding policy is optimal for the problem $\Psi(\mathbf{r}, M)$. Denoting such a policy as $\phi^*(\mathbf{r}, M)$, the proposition is proven.

\section{Proof of Lemma~\ref{lemma:alpha_tau}}
\label{sec:alpha_tau_proof}

We next prove the Lemma for $\alpha^{\tau}_j\big(j, \omega,\rho\big)$.

\begin{figure}
\centering
\includegraphics[width=3.2in]{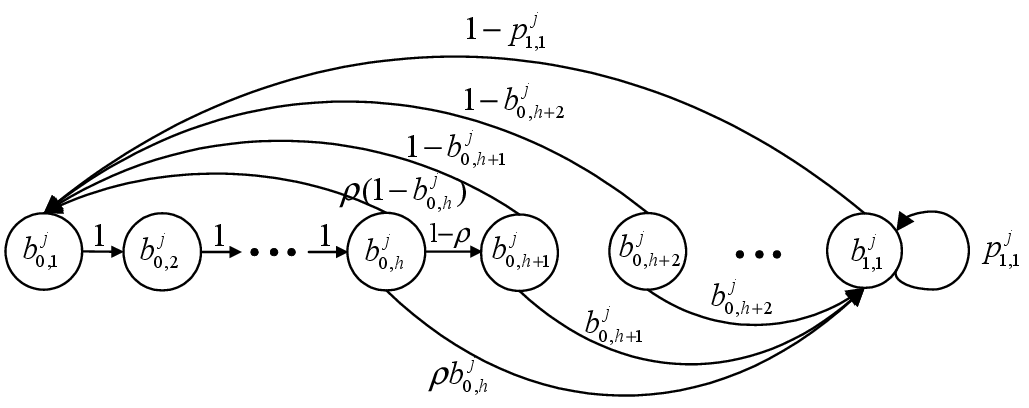}
\vspace{-4pt}
\caption{Belief value transition in steady state when $\omega{=}W_j^{\mathbf{r}}(b^j_{0,h})$.}
\vspace{-3pt}
\label{fig:belief_pos}
\end{figure}

Case (1). First consider $\alpha^{\tau}_j\big(j, W_j^{\mathbf{r}}(b^j_{0,h}),\rho\big)$ with $h<\tau$. Hence user $j$ is scheduled if its belief value is above $b^j_{0,h}$, or is scheduled with probability $\rho$ at belief value $b^j_{0,h}$. According to the belief value evolution rule~(\ref{eq:evolve}), in the next slot, its belief value will either be $p^j_{11}$ or $p^j_{01}$, depending on the whether the revealed channel state is `$0$' or `$1$' at the end of the current slot. If the user's belief value is below $b^j_{0,h}$, it will not be scheduled and its belief value will move one step toward $b^j_{0,h+1}$. Hence, in this case, the belief value evolution for user $j$ follows a Markov Chain over $\mathcal{B}^{\tau}_j$, as depicted in Fig.~\ref{fig:belief_pos}.

From Fig.~\ref{fig:belief_pos}, one can observe that the belief Markov chain is ergodic and the recurrent states are $\{b^j_{1,1},b^j_{0,l},l=1,\cdots,h+1\}$. We denote the stationary probability of belief value being $\pi_j$ as $\zeta_j(\pi_j), \pi_j\in \mathcal{B}^{\tau}_j$. The global balance equations are
\begin{align}
\rho(1{-}b^j_{0,h})\zeta_j(b^j_{0,h}){+}\zeta_j(b^j_{0,h{+}1})(1{-}b^j_{0,h{+}1})&\nonumber\\
{+}b^j_{1,1}(1{-}&p^j_{11})=\zeta_j(b^j_{0,1})\nonumber\\
\zeta_j(b^j_{0,1})=\zeta_j(b^j_{0,2})=,\cdots,=\zeta_j&(b^j_{0,h})\nonumber\\
(1-\rho)\zeta_j(b^j_{0,h})=\zeta_j(b^j_{0,h+1}&)\nonumber\\
\rho\zeta_j(b^j_{0,h})+\zeta_j(b^j_{0,h+1})=(1-p^j_{11})&\zeta_j(b^j_{1,1})\nonumber
\end{align}

From the balance equations, we can calculate the expression of the stationary probability as follows,
\begin{align}
&\zeta_j(\pi_j)\nonumber\\
\nonumber={\hspace{-5pt}}&\begin{cases}&\frac{1-p^j_{11}}{\rho b^j_{0,h}{+}(1{-}\rho)b^j_{0,h\hspace{-1pt}{+}\hspace{-1pt}1}{+}(1\hspace{-1pt}{-}\hspace{-1pt}p^j_{11})(h{+}1{-}\rho)} \text{if $\pi_j{=}b^j_{0,k}$, $k\leq h$;}\\
&\frac{(1-\rho)(1-p^j_{11})}{\rho b^j_{0,h}{+}(1{-}\rho)b^j_{0,h\hspace{-1pt}{+}\hspace{-1pt}1}{+}(1\hspace{-1pt}{-}\hspace{-1pt}p^j_{11})(h{+}1{-}\rho)} \text{if $\pi_j{=}b^j_{0,h+1}$;}\\
&\frac{b^j_{0,h+1}+\rho(b^j_{0,h}-b^j_{0,h+1})}{\rho b^j_{0,h}{+}(1{-}\rho)b^j_{0,h\hspace{-1pt}{+}\hspace{-1pt}1}{+}(1\hspace{-1pt}{-}\hspace{-1pt}p^j_{11})(h{+}1{-}\rho)} \text{if $\pi_j{=}b^j_{1,1}$;}\\
&0\text{\hspace{1.77in}otherwise.}
\end{cases}
\end{align}

Hence, the expected fraction of time transmitting to user $j$ is
\begin{align}
\alpha^{\tau}_j\big(j, W_j^{\mathbf{r}}(b^j_{0,h}),\rho\big)&=\rho\zeta_j(b^j_{0,h})+\zeta_j(b^j_{0,h+1})+\zeta_j(b^j_{1,1})\nonumber\\
&=\frac{\rho(b^j_{0,h}-b^j_{0,h+1})+1-p^j_{11}+b^j_{0,h{+}1}}{\rho b^j_{0,h}{+}(1{-}\rho)b^j_{0,h\hspace{-1pt}{+}\hspace{-1pt}1}{+}(1\hspace{-1pt}{-}\hspace{-1pt}p^j_{11})(h{+}1{-}\rho)},\nonumber
\end{align}
as given in Lemma~\ref{lemma:alpha_tau}(i). To prove part (ii), we consider its reciprocal, i.e.,
\begin{align}
&[\alpha^{\tau}_j\big(j, W_j^{\mathbf{r}}(b^j_{0,h}),\rho\big)]^{-1}\nonumber\\
=&1+\frac{(1-p^j_{11})(h-\rho)}{\rho(b^j_{0,h}-b^j_{0,h+1})+1-p^j_{11}+b^j_{0,h{+}1}} \nonumber \\
=&1{+}\frac{1-p^j_{11}}{b_{0,h{+}1}^j{-}b_{0,h}^j} \Big[1{-}\frac{1-p^j_{11}+b_{0,h+1}^j+h(b_{0,h}^j-b_{0,h+1}^j)}{\rho(b_{0,h}^j{-}b_{0,h{+}1}^j)+b_{0,h+1}^j+(1-p^j_{11})}\Big].\label{eq:beta}
\end{align}

Considering the numerator inside the parenthesis of~(\ref{eq:beta}), we have
\begin{align}
&1-p^j_{11}+b_{0,h+1}^j+h(b_{0,h}^j-b_{0,h+1}^j)\nonumber\\
\geq& 1-p^j_{11}+b_{0,h+1}^j+(h+1)(b_{0,h}^j-b_{0,h+1}^j)
\geq0,\nonumber
\end{align}
where the last inequality is from (\ref{eq:denom}). Noting that the denominator inside the parenthesis of~(\ref{eq:beta}) strictly decreases with $\rho$, hence $[\alpha^{\tau}_j\big(j, W_j^{\mathbf{r}}(b^j_{0,h}),\rho\big)]^{-1}$ strictly decreases with $\rho$. Therefore $\alpha^{\tau}_j\big(j, W_j^{\mathbf{r}}(\pi_j),\rho\big)$ strictly decreases with $\rho$ for $\pi_j$ for $\pi_j\in\{b^j_{0,1},b^j_{0,2},\cdots,b^j_{0,\tau-1}\}$

Since for $h{+}1{<}\tau$, $\alpha^{\tau}_j(j, W_j^{\mathbf{r}}(b^j_{0,h}),0)=\alpha^{\tau}_j(j, W_j^{\mathbf{r}}(b^j_{0,h+1}),1)$, we have,
\begin{align}
\alpha^{\tau}_j(j, W_j^{\mathbf{r}}(b^j_{0,h+1}),\rho)&\leq\alpha^{\tau}_j(j,W_j^{\mathbf{r}}(b^j_{0,h+1}),1)\nonumber\\
&=\alpha^{\tau}_j(j,W_j^{\mathbf{r}}(b^j_{0,h}),0)\nonumber\\
&\leq \alpha^{\tau}_j(j,W_j^{\mathbf{r}}(b^j_{0,h}),\rho).\nonumber
\end{align}

Therefore, for fixed $\rho$, ${\alpha}^{\tau}_j(j,W_j^{\mathbf{r}}(\pi_j),\rho)$ strictly decreases with $\pi_j$ for $\pi_j\in\{b^j_{0,1},b^j_{0,2},\cdots,b^j_{0,\tau-1}\}$.

Case (2). Next consider $\alpha^{\tau}_j\big(j, W_j^{\mathbf{r}}(b^j_{\tau}),\rho\big)$.  We can perform a similar analysis as in case (1) to obtain
\begin{align}
\alpha^{\tau}_j\big(j, W_j^{\mathbf{r}}(b^j_{0,\tau}),\rho\big)&=\frac{\rho(b^j_{0,\tau}-b^j_{s})+1-p^j_{11}+b^j_{s}}{\rho b^j_{0,\tau}{+}(1{-}\rho)b^j_{s}{+}(1\hspace{-1pt}{-}\hspace{-1pt}p^j_{11})(\tau{+}1{-}\rho)},\nonumber\\
[\alpha^{\tau}_j\big(j, W_j^{\mathbf{r}}(b^j_{0,\tau}),\rho\big)]^{-1}\hspace{-10pt}&\nonumber\\
&\hspace{-1in}=1{+}\frac{1-p^j_{11}}{b_{s}^j{-}b_{0,\tau}^j} \Big[1{-}\frac{1-p^j_{11}+b_{s}^j+\tau(b_{0,\tau}^j-b_{s}^j)}{\rho(b_{0,\tau}^j{-}b_{s}^j)+b_{s}^j+(1-p^j_{11})}\Big].\label{eq:betac}
\end{align}

When $\tau>\tau_0$, it can be derived that the numerator $1-p^j_{11}+b_{s}^j+\tau(b_{0,\tau}^j-b_{s}^j)$ inside (\ref{eq:betac}) is positive. Therefore $\alpha^{\tau}_j\big(j, W_j^{\mathbf{r}}(b^j_{0,\tau}),\rho\big)$ strictly increases with $\rho$. Similar to Case (1), we have
\begin{align}
\alpha^{\tau}_j(j,W_j^{\mathbf{r}}(b^j_{0,\tau}),\rho)&\leq\alpha^{\tau}_j(j,W_j^{\mathbf{r}}(b^j_{0,\tau}),1)=\alpha^{\tau}_j(j,W_j^{\mathbf{r}}(b^j_{0,\tau-1}),0)\nonumber\\
&\leq \alpha^{\tau}_j(j,W_j^{\mathbf{r}}(b^j_{0,\tau-1}),\rho).\nonumber
\end{align}

Case (3). Consider $\alpha^{\tau}_j\big(j, W_j^{\mathbf{r}}(b^j_{s}),\rho\big)$. Similar to Case (1), we obtain,
\begin{align}
\alpha^{\tau}_j\big(j, W_j^{\mathbf{r}}(b^j_{s}),\rho\big)&{=}\frac{\rho(1-p^j_{11}+b^j_{s})}{(1+\tau\rho)(1-p^j_{11})+\rho b^j_{s}}.\nonumber
\end{align}

Taking the reciprocal we have
\begin{align}
[\alpha^{\tau}_j\big(j, W_j^{\mathbf{r}}(b^j_{s}),\rho\big)]^{-1}&=\frac{1}{1-p^j_{11}+b^j_{s}}\Big((1-p^j_{11})(\tau+\frac{1}{\rho})+b^j_{s}\Big),\nonumber
\end{align}
which strictly decreases with $\rho$. Hence $\alpha^{\tau}_j\big(j, W_j^{\mathbf{r}}(b^j_{s}),\rho\big)$ strictly increases with $\rho$. We also have
\begin{align}
\alpha^{\tau}_j(j,W_j^{\mathbf{r}}(b^j_{s}),\rho)&\leq\alpha^{\tau}_j(j,W_j^{\mathbf{r}}(b^j_{s}),1)=\alpha^{\tau}_j(j,W_j^{\mathbf{r}}(b^j_{0,\tau}),0)\nonumber\\
&\leq \alpha^{\tau}_j(j,W_j^{\mathbf{r}}(b^j_{0,\tau}),\rho).\nonumber
\end{align}

From Case (1)-(3), the lemma is established for $\alpha^{\tau}_j\big(j, W_j^{\mathbf{r}}(b^j_{0,h}),\rho\big)$. Noting that for user $i\neq j$, there is no randomization associated with scheduling. Hence, the above derivation for $\alpha^{\tau}_j\big(j, W_j^{\mathbf{r}}(b^j_{0,h}),\rho\big)$ naturally extends to $\alpha^{\tau}_i\big(j, W_i^{\mathbf{r}}(b^j_{0,h}),\rho\big)$. The only change is there is no longer randomization involved. Details are hence neglected here.

\section{Proof of Lemma~\ref{lemma:eps_bound_tau}}
\label{appen:f_tau}

\subsection{Proof outline}

We establish the proof by first proving lemma~\ref{lemma:sublinear} that bounds the difference of weighted sum-throughput between policies with different threshold parameters, with respective to the difference between expected fraction of transmission time to each user. We then prove the lemma under two cases, i.e., whether $\omega^*<W^{\mathbf{r}}_i(b^i_{0,\tau})$ for all user $i$. The first case is uncomplicated to prove. For the second case, we first prove a useful fact that only one of the three cases holds: $\omega_{\tau}>\omega^*$, or $\omega_{\tau}=\omega^*$ with $\rho_{\tau}<\rho^*$ and $i_{\tau}=i^*$, or $\omega_{\tau}=\omega^*$ with $i_{\tau}>i^*$. Based on these cases, we can bound the difference between expected fraction of time transmitting to different users. We then use Lemma~\ref{lemma:sublinear} to finish the proof.

\subsection{Notations}

Recall that, in the untruncated state space, the optimal policy $\phi^*(\mathbf{r}, M)$ corresponds to the parameters $(i^*,\omega^*,\rho^*)$. Also recall that, in the truncated state space, the policy $\phi_{\tau}(\mathbf{r},M)$ corresponds to the parameter $(i_{\tau},\omega_{\tau},\rho_{\tau})$.

Over the actual \emph{untruncated} model, consider the following policy denoted as $\phi^{untrunc}_{j,\omega, \rho}$ with the parameters $(j,\omega,\rho)$: User $i$ is scheduled if $W^{\mathbf{r}}_i(\pi_i[t]){>}\omega$, or if $W^{\mathbf{r}}_i(\pi_i[t]){=}\omega^*$ with $i{>}j$. User $i$ stays idle if $W^{\mathbf{r}}_i(\pi_i^{\tau}[t]){<}\omega$, or if $W^{\mathbf{r}}_i(\pi_i^{\tau}[t]){=}\omega^*$ with $i{<}j$.
If $W^{\mathbf{r}}_i(\pi_i^{\tau}[t]){=}\omega$ with $i{=}j$, it is scheduled with probability $\rho$. In this model, similar to~(\ref{eq:alpha_tau_def}), we let ${\alpha}_i(j,\omega, \rho)$ denote the long-term expected fraction of time transmitting to user $i$ under policy $\phi^{untrunc}_{j,\omega, \rho}$, i.e.,
\begin{align}
\label{eq:alpha_def}
{\alpha}_i(j,\omega, \rho)=\limsup_{T\rightarrow \infty} \frac{1}{T} \mathbb{E}\Big[ \sum_{t=0}^{T-1} a_i^{\phi^{untrunc}_{j,\omega, \rho}}[t]\Big].
\end{align}

The closed-form expression of ${\alpha}_i(j,\omega, \rho)$ can be calculated from the same technique we used to prove Lemma~\ref{lemma:alpha_tau} as follows.
\begin{align}
\label{eq:act_time2j}
&{\alpha}_i(j,\omega, \rho)\nonumber\\
={\hspace{-5pt}}&\begin{cases}
\frac{\rho(b^i_{0,h}-b^i_{0,h+1})+1-p^i_{11}+b^i_{0,h{+}1}}{\rho b^i_{0,h}{+}(1{-}\rho)b^i_{0,h\hspace{-1pt}{+}\hspace{-1pt}1}{+}(1\hspace{-1pt}{-}\hspace{-1pt}p^i_{11})(h{+}1{-}\rho)} &\text{if $\omega{=}W^{\mathbf{r}}_i(b^i_{0,h}), i{=}j$;}\\
\frac{1-p^i_{11}+b^i_{0,h{+}1}}{b^i_{0,h{+}1}{+}(1{-}p^i_{11})(h{+}1)}&\text{if $\omega{=}W^{\mathbf{r}}_i(b^i_{0,h})$, $i{<}j$}\\
\frac{1-p^i_{11}+b^i_{0,h}}{b^i_{0,h}{+}(1{-}p^i_{11})h}&\text{if $\omega{=}W^{\mathbf{r}}_i(b^i_{0,h})$, $i>j$;}\\
&\text{\hspace{-0.95in}or if $W^{\mathbf{r}}_i(b^i_{0,h{-}1}){<}\omega{<}W^{\mathbf{r}}_i(b^i_{0,h})$, $i\neq j$}\\
0 &\text{if $\omega{\geq}W^{\mathbf{r}}_i(b^i_s)$.}
\end{cases}
\end{align}

We also let $\upsilon_i(j,\omega,\rho)$ denote the long-term expected transmission rate to user $i$, i.e.,
\begin{align}
\label{eq:upsilon}
\upsilon_i(j,\omega,\rho)=\liminf_{T\rightarrow \infty} \frac{1}{T} \mathbb{E}\Big[ \sum_{t=0}^{T-1}r_i\cdot \pi_i[t]\cdot a_i^{\phi^{untrunc}_{j,\omega, \rho}}[t]\Big],
\end{align}

Over the \emph{truncated} model, correspondingly, we let $\upsilon_i^{\tau}(j,\omega,\rho)$ denote the long-term expected transmission rate to user $i$ under policy $\phi^{trunc}_{j,\omega, \rho}$ defined in section~\ref{sec:thr_approx}, i.e.,
\begin{align}
\label{eq:upsilon_tau}
\upsilon_i^{\tau}(j,\omega,\rho)=\liminf_{T\rightarrow \infty} \frac{1}{T} \mathbb{E}\Big[ \sum_{t=0}^{T-1} r_i\cdot\pi_i^{\tau}[t]\cdot a_i^{\phi^{trunc}_{j,\omega,\rho}}[t]\Big].
\end{align}

Using techniques similar to the proof of Lemma~\ref{lemma:alpha_tau}, we can derive the analytical expressions of $\upsilon_i(j,\omega,\rho)$ and $\upsilon_i^{\tau}(j,\omega,\rho)$ as follows,
{\singlespace\begin{align}
\label{eq:act_rew_untrun}
&{\upsilon}_i(j,\omega, \rho)=\nonumber\\
&\hspace{-3pt}\begin{cases}
r_i {\cdot} \frac{\rho b_{0,h}^i+(1-\rho)b_{0,h+1}^i}{\rho b_{0,h}^i{+}(1{-}\rho)b_{0,h{+}1}^i{+}(1{-}p^i_{11})(h{+}1{-}\rho)} &\text{\hspace{-5pt}if $\omega{=}W_i^{\mathbf{r}}(b^i_{0,h})$, $i{=}j$} \\
r_i {\cdot} \frac{b_{0,h+1}^i}{b_{0,h{+}1}^i{+}(1{-}p^i_{11})(h{+}1)} &\text{\hspace{-5pt}if $\omega{=}W^{\mathbf{r}}_i(b^i_{0,h})$, $i{<}j$} \\
r_i {\cdot} \frac{b_{0,h}^i}{b_{0,h}^i{+}(1{-}p^i_{11})h} &\text{\hspace{-5pt}if $\omega{=}W_i^{\mathbf{r}}(b^i_{0,h})$, $i{>}j$} \\
&\text{\hspace{-1.1in}or if $W^{\mathbf{r}}_i(b^i_{0,h{-}1}){<}\omega{<}W^{\mathbf{r}}_i(b^i_{0,h})$, $i\neq j$}\\
0 &\text{\hspace{-5pt}if $\omega{\geq}W_i^{\mathbf{r}}(b^i_s)$.}
\end{cases}
\end{align}}

The expression of ${\upsilon}^{\tau}_j(j,\omega, \rho)$ is given as follows,
{\singlespace\begin{align}
\label{eq:act_rew}
&{\upsilon}^{\tau}_j(j,\omega, \rho)=\nonumber\\
&\hspace{-3pt}\begin{cases}
r_j {\cdot} \frac{\rho b_{0,h}^j+(1-\rho)b_{0,h+1}^j}{\rho b_{0,h}^j{+}(1{-}\rho)b_{0,h{+}1}^j{+}(1{-}p^j_{11})(h{+}1{-}\rho)} &\text{if $h{<}\tau$, $\omega{=}W^{\mathbf{r}}_j(b^j_{0,h})$;}\\
r_j {\cdot} \frac{\rho b_{0,\tau}^j+(1-\rho)b_{s}^j}{\rho b_{0,\tau}^j{+}(1{-}\rho)b_{s}^j{+}(1{-}p^j_{11})(\tau{+}1{-}\rho)} &\text{if $\omega{=}W_j^{\mathbf{r}}(b^j_{0,\tau})$;} \\
r_j {\cdot} \frac{\rho b^j_s}{(1+\tau\rho)(1{-}p^j_{11}){+}\rho b^j_s} &\text{if $\omega{=}W_j^{\mathbf{r}}(b^j_s)$;}\\
0 &\text{if $\omega{>}W_j^{\mathbf{r}}(b^j_s)$.}
\end{cases}
\end{align}}

The expression of ${\upsilon}^{\tau}_i(j,\omega, \rho)$, $i\neq j$ is expressed as follows.
{\singlespace\begin{align}
\label{eq:act_rew}
&{\upsilon}^{\tau}_i(j,\omega, \rho)=\nonumber\\
&\hspace{-3pt}\begin{cases}
r_i {\cdot} \frac{b_{0,h+1}^i}{b_{0,h{+}1}^i{+}(1{-}p^i_{11})(h{+}1)} &\text{if $h{<}\tau$, $\omega{=}W^{\mathbf{r}}_i(b^i_{0,h})$, $i{<}j$}\\
r_i {\cdot} \frac{b_{0,h}^i}{b_{0,h}^i{+}(1{-}p^i_{11})h} &\text{if $h{\leq}\tau$, $\omega{=}W^{\mathbf{r}}_i(b^i_{0,h})$, $i>j$;}\\
&\text{or if $h{\leq}\tau$,$W^{\mathbf{r}}_i(b^i_{0,h{-}1}){<}\omega{<}W^{\mathbf{r}}_i(b^i_{0,h})$}\\
r_i {\cdot} \frac{b^i_s}{(1+\tau)(1{-}p^i_{11}){+}b^i_s} &\text{if $\omega{=}W^{\mathbf{r}}_i(b^i_{0,\tau})$, $i<j$;}\\
&\text{or if $\omega{=}W^{\mathbf{r}}_i(b^i_s)$, $i>j$}\\
0 &\text{if $\omega{=}W^{\mathbf{r}}_i(b^i_s)$, $i<j$}\\
&\text{or if $\omega{>}W^{\mathbf{r}}_i(b^i_s)$.}
\end{cases}
\end{align}}

\subsection{Proof of Lemma~\ref{lemma:eps_bound_tau}}

We first prove the following lemma that provides properties of ${\alpha}^{\tau}_i(j,\omega,\rho)$ and ${\upsilon}^{\tau}_i(j,\omega,\rho)$.

\begin{lemma}
\label{lemma:sublinear}
For a user $i$, if $\tau\geq \tau_0$, we have

\noindent(i) For fixed $\pi_j{\in}\{b^j_{0,1},b^j_{0,2},{\cdots},b^j_{0,\tau},b^j_{s}\}$, ${\upsilon}^{\tau}_j(j,W_j^{\mathbf{r}}(\pi_i),\rho)$ strictly increases with $\rho$. For fixed $\rho$, ${\upsilon}^{\tau}_i(j,W_i^{\mathbf{r}}(\pi_i),\rho)$ strictly decreases with $\pi_i$ for $\pi_i\in\{b^i_{0,1},b^i_{0,2},\cdots,b^i_{0,\tau},b^i_{s}\}$ and all $i$;

\noindent(ii) for any two sets of parameter $\{j_1, \omega_1, \rho_1\}$ and $\{j_2, \omega_2, \rho_2\}$,
\begin{align}
&\Big|\hspace{2pt}{\upsilon}^{\tau}_i(j_1, \omega_1, \rho_1)-{\upsilon}^{\tau}_i(j_2, \omega_2, \rho_2) \Big|\nonumber\\
\leq &r_i \cdot \Big|\hspace{2pt}{\alpha}^{\tau}_i(j_1, \omega_1, \rho_1)-{\alpha}^{\tau}_i(j_2, \omega_2, \rho_2)\Big|.\nonumber
\end{align}
\end{lemma}

\noindent \textbf{Proof:} See Appendix~\ref{sec:sublinear_proof}. $\hfill \blacksquare$

\vspace{5pt}Note that we need $\tau\geq\tau_0$ for the proof to hold. Since the untruncated state space is in the asymptotic regime of the truncated scenario when $\tau{\rightarrow}\infty$, a straightforward extension of properties of ${\alpha}^{\tau}_i(j,\omega, \rho)$ and ${\upsilon}^{\tau}_i(j,\omega, \rho)$ in  Lemma~\ref{lemma:alpha_tau} and Lemma~\ref{lemma:sublinear} to ${\alpha}_i(j,\omega, \rho)$ and ${\upsilon}_i(j,\omega, \rho)$ in the untruncated scenario leads to the next Lemma.

\begin{lemma}
\label{lemma:sublinear2}
For a user $i$, if $\tau\geq \tau_0$, we have

\noindent(i) For fixed $\pi_j{\in}\{b^j_{0,1},b^j_{0,2},{\cdots},b^j_{0,\tau},b^j_{s}\}$, ${\upsilon}_j(j,W_j^{\mathbf{r}}(\pi_i),\rho)$ and ${\alpha}_i(j,W_i^{\mathbf{r}}(\pi_i), \rho)$ strictly increase with $\rho$. For fixed $\rho$, ${\upsilon}_i(j,W_i^{\mathbf{r}}(\pi_i),\rho)$ and ${\alpha}_i(j,W_i^{\mathbf{r}}(\pi_i), \rho)$ strictly decrease with $\pi_i$ for $\pi_i\in\{b^i_{0,1},b^i_{0,2},\cdots,b^i_{0,\tau},b^i_{s}\}$;

\noindent(ii) for any two sets of parameters $\{j_1, \omega_1, \rho_1\}$ and $\{j_2, \omega_2, \rho_2\}$,
\begin{align}
&\Big|\hspace{2pt}{\upsilon}_i(j_1,\omega_1, \rho_1){-}{\upsilon}_i(j_2,\omega_2, \rho_2) \Big|\nonumber\\
{\leq}& r_i {\cdot} \Big|\hspace{2pt}{\alpha}_i(j_1,\omega_1, \rho_1)-{\alpha}_i(j_2,\omega_2, \rho_2)\Big|.\nonumber
\end{align}
\end{lemma}

\vspace{8pt}We proceed to prove Lemma~\ref{lemma:eps_bound_tau} under two cases.

\vspace{8pt}\noindent Case (1). If the threshold $\omega^*$ satisfies $\omega^*<W^{\mathbf{r}}_i(b^i_{0,\tau})$ for all user $i$, then the approximation parameters $i_{\tau}=i^*$, $\omega_{\tau}=\omega^*$ and $\rho_{\tau}=\rho^*$. This is because, if $\omega^*<W^{\mathbf{r}}_i(b^i_{0,\tau})$ for all user $i$, no user will stay idle for more than $\tau$ slots under the optimal policy $\phi^*(\mathbf{r}, M)$. To see this in more detail, the expected amount of transmissions equals to $M$, i.e., $\sum_{i=1}^N {\alpha}^{\tau}_i(j,\omega, \rho)=M$, when $j=i^*$, $\omega=\omega^*, \rho=\rho^*$, which meets the constraint~(\ref{eq:tau_constr}). Therefore, thanks to the strict monotonicity property in Lemma~\ref{lemma:alpha_tau}(ii), the algorithm $G^{\tau}(\mathbf{r}, M)$ outputs $i_{\tau}=i^*$, $\omega_{\tau}=\omega^*$ and $\rho_{\tau}=\rho^*$, and hence policy $\phi_{\tau}(\mathbf{r}, M)$ is equivalent to the policy $\phi^*(\mathbf{r}, M)$. We hence have $\big|V^*(\mathbf{r},M){-}V_{\tau}(\mathbf{r},M)\big|{=}0$ and $Z_{\tau}(\mathbf{r},M){=}M$.

\vspace{8pt}\noindent Case (2). If there exists a user $i$ with $\omega^* \geq W^{\mathbf{r}}_i(b^i_{0,\tau})$, we let $\Theta$ denote the corresponding set of users, i.e., $\Theta=\{i: W^{\mathbf{r}}_i(b^i_{0,\tau})\leq\omega^* \}$. Therefore,
\begin{align}
&\big|V^*(\mathbf{r},M)-V_{\tau}(\mathbf{r},M)\big|\nonumber\\
=&\big| \sum_{i=1}^N \upsilon_i(i^*,\omega^*,\rho^*)- \sum_{i=1}^N {\upsilon}_i(i_{\tau},\omega_{\tau},\rho_{\tau})\big| \nonumber\\
\leq& \sum_{i\in \Theta} \Big|\upsilon_i(i^*,\omega^*,\rho^*){-}{\upsilon}_i(i_{\tau},\omega_{\tau},\rho_{\tau})\Big|\nonumber\\
&\hspace{0.4in}{+} \sum_{i \notin\Theta} \Big|\upsilon_i(i^*,\omega^*,\rho^*){-}{\upsilon}_i(i_{\tau},\omega_{\tau},\rho_{\tau})\Big|. \label{eq:V_Vtau}
\end{align}

Before bounding~(\ref{eq:V_Vtau}), we first show that, for this case, we have only one of the three cases: $\omega_{\tau}>\omega^*$, or $\omega_{\tau}=\omega^*$ with $\rho_{\tau}<\rho^*$ and $i_{\tau}=i^*$, or $\omega_{\tau}=\omega^*$ with $i_{\tau}>i^*$.

We prove the above statement by first showing that $\sum_{i=1}^N {\alpha}^{\tau}_i(i^*,\omega^*,\rho^*){\geq}\sum_{i=1}^N \alpha_i(i^*,\omega^*,\rho^*)=M$: For any user $i \notin \Theta$, we have ${\alpha}^{\tau}_i(i^*,\omega^*,\rho^*)=\alpha_i(i^*,\omega^*,\rho^*)$ since $(i^*,\omega^*,\rho^*)$ does not exceed the truncation level. For user $i \in \Theta$, 1) if $\omega^*\geq W^{\mathbf{r}}_i(b^i_{s})$, we have ${\alpha}^{\tau}_i(i^*,\omega^*,\rho^*)\geq\alpha_i(i^*,\omega^*,\rho^*)$ since $\alpha_i(i^*,\omega^*,\rho^*){=}0$. 2) If $W^{\mathbf{r}}_i(b^i_{0,\tau})<\omega^*<W^{\mathbf{r}}_i(b^i_{s})$ for $i \in \Theta$, we have
\begin{align}
\nonumber{\alpha}^{\tau}_i(i^*,\omega^*,\rho^*)=&{\alpha}^{\tau}_i(i^*,W^{\mathbf{r}}_i(b^i_{s}),1)=\frac{1-p^i_{11}+b^i_{s}}{(1+\tau)(1-p^i_{11})+b^i_{s}}\nonumber\\
>&\frac{1-p^i_{11}+b^i_{0,\tau+1}}{(1+\tau)(1-p^i_{11})+b^i_{0,\tau+1}}\nonumber\\
{=}&\alpha_i(i^*,W^{\mathbf{r}}_i(b^i_{0,\tau}),0)\geq\alpha_i(i^*,\omega^*,\rho^*),\nonumber
\end{align}
where the first equality holds because, when $W^{\mathbf{r}}_i(b^i_{0,\tau})<\omega^*<W^{\mathbf{r}}_i(b^i_{s})$, the user is scheduled when its belief value is not below $b^i_{s}$ and stays idle otherwise. Because of the truncation, the next belief value above $b^i_{0,\tau}$ is $b^i_{s}$. Since user $i^{th}$ index value will not be exactly $\omega^*$, the randomization factor $\rho^*$ at the threshold does not play a role. Hence the expected fraction of transmission time ${\alpha}^{\tau}_i(i^*,\omega^*,\rho^*)$ equals ${\alpha}^{\tau}_i(i^*,W^{\mathbf{r}}_i(b^i_{s}),1)$, i.e., transmit to user $i$ when its belief value is not below $b^i_{s}$ with probability $1$. The second and the third equality are from lemma~\ref{lemma:alpha_tau}(i) and (\ref{eq:act_time2j}), respectively. The first inequality holds since $b^i_{s}>b^i_{0,\tau+1}$. The last inequality holds because $W^{\mathbf{r}}_i(b^i_{0,\tau}){<}\omega^*{<}W^{\mathbf{r}}_i(b^i_{s})$, hence from~(\ref{eq:act_time2j}) and the monotonicity property in Lemma~\ref{lemma:sublinear2}(i),
\begin{align}
&\alpha_i(i^*,W^{\mathbf{r}}_i(b^i_{0,\tau}),0){=}\alpha_i(i^*,W^{\mathbf{r}}_i(b^i_{0,\tau+1}),1)\nonumber\\
{=}&\alpha_i(i^*,\omega^*,\rho^*) \text{\quad if $W^{\mathbf{r}}_i(b^i_{0,\tau}){<}\omega^*{<}W^{\mathbf{r}}_i(b^i_{0,\tau+1})$,}\nonumber\\
&\alpha_i(i^*,W^{\mathbf{r}}_i(b^i_{0,\tau}),0){=}\alpha_i(i^*,W^{\mathbf{r}}_i(b^i_{0,\tau+1}),1)\nonumber\\
{\geq}&\alpha_i(i^*,\omega^*,1)\geq \alpha_i(i^*,\omega^*,\rho^*)\text{\quad if $W^{\mathbf{r}}_i(b^i_{0,\tau+1}){\leq}\omega^*{<}W^{\mathbf{r}}_i(b^i_{s})$.}\nonumber
\end{align}
3) If $\omega^*=W^{\mathbf{r}}_i(b^i_{0,\tau})$, similarly, for $i \in \Theta$,
\begin{align}
\nonumber&{\alpha}^{\tau}_i(i^*,\omega^*,\rho^*)>{\alpha}_i(i^*,\omega^*,\rho^*).\nonumber \end{align}

Hence from 1)-3) we have ${\alpha}_i^{\tau}(i^*,\omega^*,\rho^*)\geq{\alpha}_i(i^*,\omega^*,\rho^*)$ for $i\in\Theta$. Also noting that, for $i\notin\Theta$, ${\alpha}_i(i^*,\omega^*,\rho^*)={\alpha}_i^{\tau}(i^*,\omega^*,\rho^*)$, we hence have
\begin{align}
&\sum_{i=1}^N {\alpha}^{\tau}_i(i^*,\omega^*,\rho^*)=\sum_{i\in\Theta} {\alpha}^{\tau}_i(i^*,\omega^*,\rho^*)+\sum_{i\notin\Theta} {\alpha}^{\tau}_i(i^*,\omega^*,\rho^*)\nonumber\\
=&\sum_{i\in\Theta} {\alpha}_i^{\tau}(i^*,\omega^*,\rho^*)+\sum_{i\notin\Theta} {\alpha}_i(i^*,\omega^*,\rho^*)\nonumber\\
\geq&\sum_{i\in\Theta} {\alpha}_i(i^*,\omega^*,\rho^*)+\sum_{i\notin\Theta} {\alpha}_i(i^*,\omega^*,\rho^*)\nonumber\\
=&\sum_{i=1}^N {\alpha}_i(i^*,\omega^*,\rho^*){=}M.\nonumber
\end{align}

\vspace{5pt}Hence if we implement the policy with threshold parameters $(i^*,\omega^*,\rho^*)$ over the fictitious truncated belief space, the expected number of transmissions will equal to or exceed the constraint. Therefore, from the monotonicity property in Lemma~\ref{lemma:alpha_tau}, to ensure the constraint~(\ref{eq:tau_constr}) on the long-term expected number of transmissions over the truncated state space, it must be one of the following three cases $\omega_{\tau}>\omega^*$, or $\omega_{\tau}=\omega^*$ with $\rho_{\tau}<\rho^*$ and $i_{\tau}=i^*$, or $\omega_{\tau}=\omega^*$ with $i_{\tau}>i^*$. From this property as well as Lemma~\ref{lemma:sublinear2}(i), we have,
\begin{align}
{\alpha}_i(i_{\tau},\omega_{\tau},\rho_{\tau})\leq {\alpha}_i(i^*,\omega^*,\rho^*) \text{\ for all $i$,} \label{eq:alpha_relation}
\end{align}
and, because $i\in\Theta$,
\begin{align}
\upsilon_i(i_{\tau},\omega_{\tau},\rho_{\tau}) &{\leq} \upsilon_i(i^*,\omega^*,\rho^*){\leq} \upsilon_i(i,W_i^{\mathbf{r}}(b^i_{0,\tau}),1), \text{for $i{\in}\Theta$.}\label{eq:upsilon}
\end{align}

Hence, for $i\in\Theta$,
\begin{align}
\big|\upsilon_i(i^*,\omega^*,\rho^*){-}{\upsilon}_i(i_{\tau},\omega_{\tau},\rho_{\tau})\big|&{\leq} {\upsilon}_i(i,W_i^{\mathbf{r}}(b^i_{0,\tau}),1)\nonumber\\
&{\leq} r_i {\cdot} {\alpha}_i(i,W_i^{\mathbf{r}}(b^i_{0,\tau}),1), \label{eq:v_r_diff_vin}
\end{align}
where the first inequality is from~(\ref{eq:upsilon}) and the last equality holds because instantaneous reward is upper bounded by $r_i$.

Similar to (\ref{eq:upsilon}), from the monotonicity properties of ${\alpha}^{\tau}_i(j,\omega,\rho)$ and ${\alpha}_i(j,\omega,\rho)$ and because $i\in\Theta$,
\begin{align}
{\alpha}^{\tau}_i(i_{\tau},\omega_{\tau},\rho_{\tau})&\leq {\alpha}^{\tau}_i(i^*,\omega^*,\rho^*) \leq {\alpha}_i^{\tau}(i,W_i^{\mathbf{r}}(b^i_{0,\tau}),1) \text{, $i\in\Theta$,}\label{eq:alpha}\\
\alpha_i(i^*,\omega^*,\rho^*)&\leq \alpha_i(i^*,\omega^*,1) \leq \alpha_i(i,W_i^{\mathbf{r}}(b^i_{0,\tau}),1)\text{, $i\in\Theta$.}\label{eq:alpha_bar}
\end{align}

For $i \notin \Theta$, we have ${\alpha}^{\tau}_i(i^*,\omega^*,\rho^*)=\alpha_i(i^*,\omega^*,\rho^*)$.
Hence,
\begin{align}
&\hspace{-5pt}\sum_{i \notin \Theta}\big|\upsilon_i(i^*,\omega^*,\rho^*){-}{\upsilon}_i(i_{\tau},\omega_{\tau},\rho_{\tau})\big|\nonumber\\
{\leq}&\sum_{i \notin \Theta} r_i {\cdot} \big|{\alpha}_i(i^*,\omega^*,\rho^*){-}{\alpha}_i(i_{\tau},\omega_{\tau},\rho_{\tau})\big|\nonumber \\
=&\sum_{i \notin \Theta} r_i \cdot \big[\alpha_i(i^*,\omega^*,\rho^*)-{\alpha}_i(i_{\tau},\omega_{\tau},\rho_{\tau})\big]\nonumber\\
\leq& \sum_{i \notin \Theta} r_i \cdot \sum_{i \notin \Theta}\big[\alpha_i(i^*,\omega^*,\rho^*)-{\alpha}_i(i_{\tau},\omega_{\tau},\rho_{\tau})\big]\nonumber
\end{align}
\begin{align}
\leq& \sum_{i \notin \Theta} r_i \cdot\Big[ \sum_{i \notin \Theta}\big[\alpha_i(i^*,\omega^*,\rho^*)-{\alpha}^{\tau}_i(i_{\tau},\omega_{\tau},\rho_{\tau})\big]+\nonumber\\
&\hspace{0.4in}\sum_{i \notin \Theta}\big[{\alpha}^{\tau}_i(i_{\tau},\omega_{\tau},\rho_{\tau})-{\alpha}_i(i_{\tau},\omega_{\tau},\rho_{\tau})\big]\Big],\label{eq:v_rs_diff_notin}
\end{align}
where the first inequality is from Lemma~\ref{lemma:sublinear2}(ii) and the first equality holds from~(\ref{eq:alpha_relation}).

Consider the first summand inside the parenthesis of (\ref{eq:v_rs_diff_notin}). Since $\sum_{i=1}^N \alpha_i(i^*,\omega^*,\rho^*)=\sum_{i=1}^N {\alpha}^{\tau}_i(i_{\tau},\omega_{\tau},\rho_{\tau})=M$, subtracting both sides by $\sum_{i \notin \Theta}{\alpha}^{\tau}_i(i_{\tau},\omega_{\tau},\rho_{\tau})+\sum_{i \in \Theta}\alpha_i(i^*,\omega^*,\rho^*)$ we have
\begin{align}
&\sum_{i \notin \Theta}\big[\alpha_i(i^*,\omega^*,\rho^*){-} {\alpha}^{\tau}_i(i_{\tau},\omega_{\tau},\rho_{\tau})\big]\nonumber\\
=&\sum_{i \in \Theta}\big[{\alpha}^{\tau}_i(i_{\tau},\omega_{\tau},\rho_{\tau}){-}\alpha_i(i^*,\omega^*,\rho^*) \big]\nonumber \\
\leq& \sum_{i \in \Theta}\big|{\alpha}^{\tau}_i(i_{\tau},\omega_{\tau},\rho_{\tau}){-}\alpha_i(i^*,\omega^*,\rho^*)\big|. \label{eq:Jineq}
\end{align}

Note that, for $i \in \Theta$, from (\ref{eq:alpha})-(\ref{eq:alpha_bar}),
\begin{align}
\big|{\alpha}^{\tau}_i(i_{\tau},\omega_{\tau},\rho_{\tau})-\alpha_i(i^*,\omega^*,\rho^*)\big|\leq
\alpha_i(i,W_i^{\mathbf{r}}(b^i_{0,\tau}),1). \label{eq:v_r_diff_ain}
\end{align}

Substituting~(\ref{eq:v_r_diff_ain}) back to~(\ref{eq:Jineq}), we have
\begin{align}
\sum_{i \notin \Theta}\big[\alpha_i(i^*,\omega^*,\rho^*){-} {\alpha}^{\tau}_i(i_{\tau},\omega_{\tau},\rho_{\tau})\big] \leq \sum_{i \in \Theta}\alpha_i(i,W_i^{\mathbf{r}}(b^i_{0,\tau}),1). \label{eq:v_r_diff_ain2}
\end{align}

Now consider the second summand inside (\ref{eq:v_rs_diff_notin}), we have, for $i\notin \Theta$,
\begin{align}
&{\alpha}^{\tau}_i(i_{\tau},\omega_{\tau},\rho_{\tau}){-}{\alpha}_i(i_{\tau},\omega_{\tau},\rho_{\tau})
=0\text{, if $\omega_{\tau}<W_i^{\mathbf{r}}(b^i_{0,\tau})$,}\label{eq:second_term_diff1}\\
&{\alpha}^{\tau}_i(i_{\tau},\omega_{\tau},\rho_{\tau}){-}{\alpha}_i(i_{\tau},\omega_{\tau},\rho_{\tau})\nonumber\\
\leq& \alpha_i(i,W_i^{\mathbf{r}}(b^i_{0,\tau}),1)\text{, if $\omega_{\tau}{=}W_i^{\mathbf{r}}(b^i_{0,\tau})$,}
\label{eq:second_term_diff2}
\end{align}
where (\ref{eq:second_term_diff2}) holds because both ${\alpha}^{\tau}_i(i_{\tau},\omega_{\tau},\rho_{\tau})\leq \alpha_i(i,W_i^{\mathbf{r}}(b^i_{0,\tau}),1)$ and ${\alpha}_i(i_{\tau},\omega_{\tau},\rho_{\tau})\leq\alpha_i(i,W_i^{\mathbf{r}}(b^i_{0,\tau}),1)$. Therefore,
\begin{align}
&\sum_{i \notin \Theta}\big[{\alpha}^{\tau}_i(i_{\tau},\omega_{\tau},\rho_{\tau}){-}{\alpha}_i(i_{\tau},\omega_{\tau},\rho_{\tau})\big]\nonumber\\
\leq& \sum_{i \notin \Theta}\alpha_i(i,W_i^{\mathbf{r}}(b^i_{0,\tau}),1). \label{eq:v_r_diff_ain3}
\end{align}

Substituting (\ref{eq:v_r_diff_ain2}) and (\ref{eq:v_r_diff_ain3}) in (\ref{eq:v_rs_diff_notin}),
\begin{align}
\sum_{i \notin \Theta}\hspace{-3pt}\big|\upsilon_i(i^*,\omega^*,\rho^*){-}{\upsilon}_i(i_{\tau},\omega_{\tau},\rho_{\tau})\big|{\leq}
\hspace{-3pt}\sum_{i \notin \Theta} r_i\hspace{-3pt}\sum_{i=1}^N {\alpha}_i(i,W_i^{\mathbf{r}}(b^i_{0,\tau}),1). \label{eq:V_noTheta}
\end{align}

From (\ref{eq:v_r_diff_vin}) and (\ref{eq:V_noTheta}), the difference in (\ref{eq:V_Vtau}) can be bounded as follows,
\begin{align}
&\big|V^*(\mathbf{r},M)-V_{\tau}^*(\mathbf{r},M)\big|\nonumber\\
\leq &\sum_{i\in \Theta} r_i \cdot  {\alpha}_i(i,W_i^{\mathbf{r}}(b^i_{0,\tau}),1)+\sum_{i \notin \Theta} r_i \sum_{i=1}^N {\alpha}_i(i,W_i^{\mathbf{r}}(b^i_{0,\tau}),1) \nonumber \\
\leq&\sum_{i=1}^N r_i \cdot \sum_{i=1}^N {\alpha}_i\big(i,W_i^{\mathbf{r}}(b^i_{0,\tau}),1\big). \nonumber
\end{align}

We let $f_i(\tau){=}{\alpha}_i\big(i,W_i^{\mathbf{r}}(b^i_{0,\tau}),1\big)$ and $f(\tau){=}\sum_{i=1}^N f_i(\tau)$. Since ${\alpha}_i\big(i,W_i^{\mathbf{r}}(b^i_{0,\tau}),1\big)\rightarrow 0$ as $\tau\rightarrow \infty$, part (i) of the lemma is established. From~(\ref{eq:alpha_relation}), we have
\begin{align}
Z_{\tau}(\mathbf{q},M)=\sum_{i=1}^N {\alpha}_i(i_{\tau},\omega_{\tau},\rho_{\tau})\leq \sum_{i=1}^N {\alpha}_i(i^*,\omega^*,\rho^*)=M,\nonumber
\end{align}
which proves part (ii). \hfill $\blacksquare$

\vspace{4pt}

\section{Proof of Proposition \ref{prop:tau_bound}}
\label{appen:thr_opt}

Define Lyapunov function $L(\mathbf{q})=\frac{1}{2}\sum_{i=1}^N q_i^2$. We consider \emph{the $T$-frame average Lyapunov drift $\Delta L(\mathbf{q}[kT])$ over the $k$-th frame}, expressed as,
\begin{align}
&\Delta L(\mathbf{q}[kT]) / T\nonumber\\
=&\frac{1}{T} \mathbb{E} \Big[L(\mathbf{q}[(k+1)T])- L(\mathbf{q}[kT]) \big | \ \mathbf{q}[kT], \bm \pi[kT] \Big]\nonumber\\
\leq & BT+ \sum_{i=1}^N q_i[kT] \cdot \lambda_i - \sum_{i=1}^N q_i[kT]\cdot\frac{1}{T} \nonumber\\
&\hspace{-8pt}\cdot\mathbb{E}\Big[ \sum_{t=0}^{T-1} \pi_i[kT{+}t] {\cdot} a_i^{\phi_{\tau}(\mathbf{q}[kT], M{-}g(\tau)/2)}[kT{+}t] \Big | \bm \pi[kT] \Big],\label{eq:drift}
\end{align}
where $B$ is a constant whose value is determined by the second moment of the arrival process \cite{Neely_tutr}. Because $\bm \lambda+g(\tau) \mathbf{1} \in \bm\Gamma$, for any non-negative vector $\mathbf{q}$, we have
\begin{align}
\sum_{i=1}^N q_i \cdot (\lambda_i + g(\tau)) \leq V^*(\mathbf{q}, M),\nonumber
\end{align}
where $V^*(\mathbf{q},M)$ is defined in (\ref{eq:thr_nontrun}).
The Lyapunov drift (\ref{eq:drift}) now becomes,
\begin{align}
\Delta L(\mathbf{q}[kT])/T&\leq BT{-}g(\tau) \sum_{i=1}^N q_i[kT]{+}\nonumber \\
&V^*(\mathbf{q}[kT],M){-} V_{\tau}^T(\mathbf{q}[kT],M{-}g(\tau)/2)\nonumber
\end{align}
\begin{align}
&= BT{-}g(\tau) \sum_{i=1}^N q_i[kT]{+} V^*(\mathbf{q}[kT],M){-}V_{\tau}(\mathbf{q}[kT],M)\nonumber \\
&+V_{\tau}(\mathbf{q}[kT],M){-}V_{\tau}(\mathbf{q}[kT],M{-}g(\tau)/2)\nonumber\\
&+V_{\tau}(\mathbf{q}[kT],M{-}g(\tau)/2){-}V_{\tau}^T(\mathbf{q}[kT],M{-}g(\tau)/2).\label{eq:drift_decomp}
\end{align}
where $V_{\tau}(\mathbf{q}[kT],M)$ is defined in (\ref{eq:thr_trun}), and $V_{\tau}^T(\mathbf{q}[kT],M)$ is the $T$-horizon expected transmission rate achieved under the policy $\phi_{\tau}(\mathbf{q}[kT], M)$, i.e.,
\begin{align}
\hspace{-6pt}&V_{\tau}^T\hspace{-2pt} (\mathbf{q}[kT],M)\nonumber\\
{=}&\sum_{i=1}^N q_i[kT]\frac{1}{T} \mathbb{E}\Big[ \sum_{t=0}^{T-1}\hspace{-2pt} \pi_i[kT{+}t] {\cdot} a_i^{\phi_{\tau}(\mathbf{q}[kT], M)}[kT{+}t] \Big | \bm \pi[kT] \Big]. \nonumber
\end{align}

Note that, in~(\ref{eq:drift_decomp}), the difference $V^*(\mathbf{q}[kT],M)-V_{\tau}(\mathbf{q}[kT],M)$ is bounded in Lemma~\ref{lemma:eps_bound_tau}. We proceed to bound the rest of the terms in~(\ref{eq:drift_decomp}). Specifically, the difference $V_{\tau}(\mathbf{q}[kT],M{-}g(\tau)/2){-}V_{\tau}^T(\mathbf{q}[kT],M{-}g(\tau)/2)$ is bounded in Lemma~\ref{lemma:exp_decay}, and the difference $V_{\tau}(\mathbf{q}[kT],M)-V_{\tau}(\mathbf{q}[kT],M{-}g(\tau)/2)$ is bounded in Lemma~\ref{lemma:R_bound}. These bounds help us to bound the Lyapunov drift $\Delta L(\mathbf{q}[kT])/T$ and later to establish the proof using Lyapunov stability theory.

We denote $Z_{\tau}^T(\mathbf{q},M)$ as the finite $T$-horizon expected number of transmissions, under the policy $\phi_{\tau}(\mathbf{q}[kT], M)$, i.e.,
\begin{align}
Z_{\tau}^T(\mathbf{q},M)&=\frac{1}{T} \mathbb{E}\Big[ \sum_{t=0}^{T-1} \sum_{i=1}^N a_i^{\phi_{\tau}(\mathbf{q},M)}[t]\Big]. \nonumber
\end{align}

The next lemma states that, as the length of the time horizon tends to infinity, the expected achieved rate in finite horizon asymptotically converges to infinite horizon achievable rate, and the expected number of transmissions converges to the value $M$.

\vspace{8pt}\begin{lemma}
\label{lemma:exp_decay}
For any M and $\kappa>0$, we have, uniformly over $\mathbf{q}$, $M$, and the initial state $\bm \pi[kT]$,

\noindent(a) there exist positive constants $c_1$ and $c_2$ such that
\begin{align}
\Big|V_{\tau}(\mathbf{q},M)-V^T_{\tau}(\mathbf{q},M)\Big| < \big(\kappa+c_1 \exp(-c_2 T) \big) \sum_{i=1}^N q_i. \nonumber
\end{align}

\noindent(b) there exist positive constants $d_1$ and $d_2$ such that
\begin{align}
\Big|Z_{\tau}^T(\mathbf{q},M)-M\Big| < \big(\kappa+d_1 \exp(-d_2 T) \big). \nonumber
\end{align}
\end{lemma}

\vspace{8pt}\noindent \textbf{Proof:}
We first prove part (a). We define the random variable $\mu_{\tau}^T(\mathbf{q},M)$ as
\begin{align}
\mu_{\tau}^T(\mathbf{q},M)=\sum_{i=1}^N q_i \frac{1}{T} \sum_{t=0}^{T-1} \pi_i[kT{+}t] \cdot a_i^{\phi_{\tau}(\mathbf{q}, M)}[kT{+}t].\nonumber
\end{align}

Therefore, $V_{\tau}^T(\mathbf{q},M)=\mathbb{E}\big[\mu_{\tau}^T(\mathbf{q},M)\big]$.
We denote event $\Omega := \big\{\big|\mu_{\tau}^T(\mathbf{q},M)-V_{\tau}(\mathbf{q},M)\big|\leq \kappa \sum_{i=1}^N q_i \big\}$, then
\begin{align}
&\mathbb{E}\Big[\big|\mu_{\tau}^T(\mathbf{q},M)-V_{\tau}(\mathbf{q},M)\big|\Big]\nonumber \\
\leq & \mathbb{E}\Big[\big|\mu_{\tau}^T(\mathbf{q},M)-V_{\tau}(\mathbf{q},M)\big|\Big| \Omega \Big]\cdot \Pr(\Omega)\nonumber\\
&\hspace{0.9in}+\mathbb{E}\Big[\big|\mu_{\tau}^T(\mathbf{q},M)-V_{\tau}(\mathbf{q},M)\big|\Big| \overline{\Omega} \Big]\cdot \Pr(\overline{\Omega})\nonumber \\
\leq& \kappa\hspace{-3pt} \sum_{i=1}^N \hspace{-1pt}q_i {+}\hspace{-3pt}\sum_{i=1}^N \hspace{-1pt}q_i {\cdot} \hspace{-1pt} \Pr\hspace{-1pt}\big(\big|\mu_{\tau}^T(\mathbf{q},M){-}V_{\tau}(\mathbf{q},M)\hspace{-2pt}\big|{>} \kappa \hspace{-4pt}\sum_{i=1}^N q_i \big).\label{eq:kappa_plus}
\end{align}

Note that
\begin{align}
&\big|\mu_{\tau}^T(\mathbf{q},M)-V_{\tau}(\mathbf{q},M)\big|\nonumber \\
=&\Big| \sum_{i=1}^N q_i \cdot \Big[\frac{1}{T} \sum_{t=0}^{T-1} \pi_i[kT+t] \cdot a_i^{\phi^{\tau}(\bm q,M)}[kT+t]\nonumber\\
&\hspace{0.5in}-\lim_{\mathbb{T} \rightarrow \infty}\frac{1}{\mathbb{T}}\sum_{t=0}^{\mathbb{T}-1} \pi_i[kT+t] \cdot a_i^{\phi^{\tau}(\bm q,M)}[kT+t]\Big] \Big|\nonumber \\
\leq & \sum_{i=1}^N q_i \cdot \Big[\sum_{i=1}^N \Big[\frac{1}{T} \sum_{t=0}^{T-1} \pi_i[kT+t] \cdot a_i^{\phi^{\tau}(\mathbf{q},M)}[kT+t]\nonumber\\
&\hspace{0.3in}-\lim_{\mathbb{T} \rightarrow \infty} \frac{1}{\mathbb{T}} \sum_{t=0}^{\mathbb{T}-1} \pi_i[kT+t] \cdot a_i^{\phi^{\tau}(\mathbf{q},M)}[kT+t]\Big]^2\Big]^{\frac{1}{2}}\nonumber\\
:=& \sum_{i=1}^N q_i {\cdot} \big\| \bm{\eta}^{\tau}(\mathbf{q},M)-\bm{\eta}^{\tau}_T(\mathbf{q},M) \big\|. \nonumber
\end{align}
where the inequality follows from Cauchy-Schwarz inequality and $\bm \eta^{\tau}(\mathbf{q},M)$ and $\bm \eta^{\tau}_{T}(\mathbf{q},M)$ are vectors with
\begin{align}
\eta^{\tau}_i(\mathbf{q},M)&=\lim_{\mathbb{T} \rightarrow \infty} \frac{1}{\mathbb{T}} \sum_{t=0}^{\mathbb{T}-1} \pi_i[kT{+}t] \cdot a_i^{\phi^{\tau}(\mathbf{q},M)}[kT{+}t], \label{eq:eta}\\
\eta^{\tau}_{T,i}(\bm q,M)&=\frac{1}{T} \sum_{t=0}^{T-1}\pi_i[t]\cdot a_i^{\phi^{\tau}(\mathbf{q},M)}[kT+t]. \label{eq:etaT}
\end{align}

Therefore,
\begin{align}
&\Pr\big(\big|\mu_{\tau}^T(\mathbf{q},M)-V_{\tau}(\mathbf{q},M)\big|> \kappa \sum_{i=1}^N q_i\big)\nonumber\\
\leq& \Pr\big(\big\| \bm{\eta}^{\tau}(\mathbf{q},M)-\bm{\eta}^{\tau}_T(\mathbf{q},M) \big\| > \kappa \big)\nonumber \\
\leq& \Pr\Big(\cup_{i=1}^N \big\{\big| \eta_{T,i}^{\tau}(\mathbf{q},M) -\eta_i(\mathbf{q},M) \big| > \kappa/N \big\}\Big)\nonumber \\
\leq& \sum_{i=1}^N \Pr\big(\big| \eta_{T,i}^{\tau}(\mathbf{q},M)-\eta^{\tau}_i(\mathbf{q},M) \big| > \kappa/N \big). \label{eq:P_kappa}
\end{align}

Recall that, under the policy $\phi^{\tau}(\mathbf{q},M)$, the belief states of different users, i.e., $\{\mathcal{B}_i^{\tau}, i=1,\cdots,N\}$, are sorted, in the initialization phase given by algorithm $G^{\tau}(\mathbf{q},M)$, in the vector $\mathbf{w}$ according to their $\mathbf{q}$-weighted index values. Consider another vector $\mathbf{\bm\varsigma}$ where each element $\varsigma_i$ corresponds to the unique belief state the $i^{th}$ element ${w}_i$ represents. So each weighing vector $\mathbf{q}$ corresponds to a vector $\mathbf{w}$ and hence $\mathbf{\bm\varsigma}$. Note that, the activation/passive scheduling decision to a user depends on the the location of the threshold for transmission, i.e., above which belief value the user is scheduled and with how much randomization. From the implementation of algorithm $G^{\tau}(\mathbf{q},M)$, as long as different policies correspond to the same $\mathbf{\bm\varsigma}$, for each user, the transmission/idle action (at each belief state) is the same function of belief state, and hence the belief state of each user evolves as the same finite-state space ergodic Markov chain. Therefore, for a policy, denoted by $\phi^{\mathbf{\bm\varsigma}}$, that corresponds to a vector $\mathbf{\bm\varsigma}$, there exist constants $c^{\mathbf{\bm\varsigma}}_1$ and $c^\mathbf{\bm\varsigma}_2$ such that, for each user $i$ uniform over the initial belief state and $\mathbf{q}$ \cite{GlynnOrmoneit} ,
\begin{align}
&\Pr\Big(\Big|\frac{1}{T} \sum_{t=0}^{T-1}\pi_i[t]\cdot a_i^{\phi^{\mathbf{\bm\varsigma}}}[t]{-}\lim_{\mathbb{T} \rightarrow \infty} \frac{1}{\mathbb{T}} \sum_{t=0}^{T-1} \pi_i[t] {\cdot} a_i^{\phi^{\mathbf{\bm\varsigma}}}[t]\Big|>\kappa/N\Big)\nonumber\\
<& \hspace{3pt}c^{\phi^{\mathbf{\bm\varsigma}}}_1\exp(-c^{\phi^{\mathbf{\bm\varsigma}}}_2 T).\label{eq:exp_conv}
\end{align}

Note that the number of users, as well as the number of vectors $\mathbf{\bm\varsigma}$, are finite. From (\ref{eq:eta})-(\ref{eq:exp_conv}), there exist constants $c_1$ and $c_2$ such that, regardless of $\mathbf{q}$ and the initial belief state,
\begin{align}
\Pr\big(\big|\mu_{\tau}^T(\mathbf{q},M)-V_{\tau}(\mathbf{q},M)\big|> \kappa \sum_{i=1}^N q_i\big)<c_1 \exp(-c_2 T). \nonumber
\end{align}

Substituting the above inequality in (\ref{eq:kappa_plus}), part(a) thus holds.

The proof of part (b) follows a similar approach as part (a). Here, the immediate reward is $a_i^{\phi^{\tau}(\mathbf{q}, M)}[kT+t]$ instead of $\pi_i[kT+t]\cdot a_i^{\phi^{\tau}(\mathbf{q}, M)}[kT+t]$. $\hfill \blacksquare$

\vspace{10pt}

\begin{lemma}\label{lemma:R_bound}
When $\tau{>}\tau_0$, for any $\epsilon{>}0$, the difference between the expected transmission rate achieved under policy $\phi_{\tau}(\mathbf{q},M)$ and $\phi_{\tau}(\mathbf{q},M-\epsilon)$ satisfies the following bound,
\begin{align}
\big|V_{\tau}(\mathbf{q},M)-V_{\tau}(\mathbf{q},M-\epsilon)\big| \leq \epsilon \sum_{i=1}^N q_i.\nonumber
\end{align}
\end{lemma}

\noindent \textbf{Proof:}
Suppose, under the weight $\bm q$, the policies $\phi_{\tau}(\bm q, M)$ and $\phi_{\tau}(\bm q, M-\epsilon)$ correspond to parameter set $\{i^{\tau}_M,\omega^{\tau}_{M},\rho^{\tau}_{M}\}$ and $(i^{\tau}_{M-\epsilon},\omega^{\tau}_{M-\epsilon},\rho^{\tau}_{M-\epsilon})$, respectively. For user $i$, we let $y_i(\epsilon)$ denote be the difference between activation time under policy $\phi_{\tau}(\bm q, M-\epsilon)$ and $\phi_{\tau}(\bm q, M)$, i.e., $y_i(\epsilon)={\alpha}_i(i^{\tau}_{M},\omega^{\tau}_{M},\rho^{\tau}_{M})-{\alpha}_i(i^{\tau}_{M-\epsilon},\omega^{\tau}_{M-\epsilon},\rho^{\tau}_{M-\epsilon})$, where, recall that, ${\alpha}_i(j,\omega,\rho)$ is defined in~(\ref{eq:act_time2j}). From Lemma~\ref{lemma:sublinear2}(i), we have $y_i(\epsilon)\geq 0, \forall i$. Since the difference of the total expected number of transmissions between the two policies is $\epsilon$, we have $\sum_{i=1}^N y_i(\epsilon)=\epsilon$. From Lemma~\ref{lemma:sublinear2}(ii), we have,
\begin{align}
&\big|V_{\tau}(\bm q,M)-V_{\tau}(\bm q,M{-}\epsilon)\big|\nonumber\\
=&\Big|\sum_{i=1}^N v_i(i^{\tau}_{M},\omega^{\tau}_{M},\rho^{\tau}_{M})- \sum_{i=1}^N {\upsilon}_i(i^{\tau}_{M{-}\epsilon},\omega^{\tau}_{M{-}\epsilon},\rho^{\tau}_{M{-}\epsilon}) \Big| \nonumber\\
\leq &\sum_{i=1}^N \Big| v_i(i^{\tau}_{M},\omega^{\tau}_{M},\rho^{\tau}_{M})-{\upsilon}_i(i^{\tau}_{M{-}\epsilon},\omega^{\tau}_{M{-}\epsilon},\rho^{\tau}_{M{-}\epsilon})\Big|\nonumber \\
\leq &\sum_{i=1}^N q_i\cdot \Big|{\alpha}_i(i^{\tau}_{M},\omega^{\tau}_{M},\rho^{\tau}_{M})-{\alpha}_i(i^{\tau}_{M{-}\epsilon},\omega^{\tau}_{M{-}\epsilon},\rho^{\tau}_{M{-}\epsilon})\Big|\nonumber \\ =& \sum_{i=1}^N q_i \cdot y_i(\epsilon)\leq \sum_{i=1}^N q_i \Big[\sum_{j=1}^N y_j(\epsilon)\Big]=\epsilon \sum_{i=1}^N q_i. \nonumber
\end{align}

We hence have proved the lemma.$\hfill \blacksquare$
\vspace{5pt}

From Lemma~\ref{lemma:eps_bound_tau} and Lemma~\ref
{lemma:exp_decay}-\ref{lemma:R_bound}, the Lyapunov drift (\ref{eq:drift_decomp}) can be further bounded as follows,
\begin{align}
&\Delta L(\mathbf{q}[kT])/ T \nonumber\\
\leq & BT{+}\nonumber\\
&\hspace{0.1in}\Big[{-}g(\tau){+}f(\tau){+}\frac{g(\tau)}{2}+\big(\kappa+c_1 \exp(-c_2 T) \big)\Big]\cdot \sum_{i=1}^N q_i[kT] \nonumber \\
=& BT{+}\Big[{-}\frac{g(\tau)}{2}{+}f(\tau){+}\big(\kappa{+}c_1 \exp({-}c_2 T) \big)\Big] \sum_{i=1}^N q_i[kT]\nonumber\\
=& BT+\Big[-f(\tau)/2+\big[\kappa+c_1 \exp(-c_2 T) \big]\Big] \sum_{i=1}^N q_i[kT] \label{eq:drift_bound}
\end{align}
where the last equality holds because we let $g(\tau)=3f(\tau)$. For fixed $\tau$, by choosing $\kappa$ sufficiently small and $T$ sufficiently large, say $T>T_1$, the Lyapunov drift is negative whenever the sum of the queue lengths gets sufficiently large. Therefore, the queues are stable according to the Foster-Lyapunov criterion.

Note that, under the policy $\text{Q-Index}_{\tau}(T,M{-}g(\tau)/2)$, the expected number of transmissions in the $k$-th frame, $Z_{\tau}^T(\mathbf{q}[kT],M-g(\tau)/2)$, is bounded by Lemma~\ref{lemma:exp_decay} as,
\begin{align}
\Big|Z_{\tau}^T(\mathbf{q}[kT],M{-}g(\tau)/2){-}(M{-}g(\tau)/2)\Big| {<} \big(\kappa{+}d_1 \exp(-d_2 T) \big), \nonumber
\end{align}
for some constant $d_1$ and $d_2$. Therefore, there exists $T_2$ such that $Z_{\tau}^T(\mathbf{q}[kT],M-g(\tau)/2)<M$ for $T>T_2$. Hence, the long term constraint on the average number of transmissions is satisfied. From Lemma~\ref{lemma:eps_bound_tau}, we have $\lim_{\tau \rightarrow \infty}g(\tau)=0$. Letting $T'=\max\{T_1, T_2\}$, the proposition is then established.

\section{Proof of Lemma~\ref{lemma:sublinear}}
\label{sec:sublinear_proof}

\noindent(i) We first prove part (i) of the lemma with $i=j$.

Case (1). If $\pi_j=b^j_{0,h}$ and $h<\tau$, we consider the reciprocal of $\upsilon_j^{\tau}(W_j^{\mathbf{r}}(b^j_{0,h}), \rho)$,
\begin{align}
&r_j\cdot [\upsilon_j^{\tau}(j,W_j^{\mathbf{r}}(b^j_{0,h}),\rho)]^{-1}= 1+\frac{(1-p^j_{11})(h+1-\rho)}{\rho (b_{0,h}^j-b_{0,h+1}^j)+b_{0,h+1}^j} \nonumber \\
=&1{+}\frac{1-p^j_{11}}{b_{0,h+1}^j{-}b_{0,h}^j} \Big[1{+}\frac{b_{0,h+1}^j{-}(h{+}1)(b_{0,h+1}^j{-}b_{0,h}^j)}{\rho(b_{0,h+1}^j-b_{0,h}^j)-b_{0,h+1}^j}\Big]\label{eq:r_i}
\end{align}

Consider the numerator in the parenthesis of~(\ref{eq:r_i})
\begin{align}
&b_{0,h+1}^j{-}(h{+}1)(b_{0,h+1}^j{-}b_{0,h}^j)=(h+1) b_{0,h}^j-h b_{0,h+1}^j\nonumber\\
=& [1+(1-p^j_{11}+p^j_{01})h]b_{0,h}^j-h p^j_{01}\nonumber\\
=&\frac{p^j_{01}[1-(p^j_{11}-p^j_{01})^h]}{1-(p^j_{11}-p^j_{01})}-h p^j_{01} (p^j_{11}-p^j_{01})^h\nonumber\\
=&p^j_{01}[1{+}(p^j_{11}{-}p^j_{01}){+}{\cdots}{+}(p^j_{11}{-}p^j_{01})^{h-1}]{-}h p^j_{01}(p^j_{11}{-}p^j_{01})^h\nonumber\\
> & h p^j_{01} (p^j_{11}{-}p^j_{01})^h-h p^j_{01} (p^j_{11}{-}p^j_{01})^h=0.\label{eq:denom}
\end{align}

Since the denominator in the parenthesis of (\ref{eq:r_i}) strictly increases with $\rho$, $[\upsilon_j^{\tau}(j,W_j^{\mathbf{r}}(b^j_{0,h}),\rho)]^{-1}$ strictly decreases with $\rho$ and hence $\upsilon_j^{\tau}(W_j^{\mathbf{r}}(j,b^j_{0,h}),\rho)$ strictly increases with $\rho$ in this case.
\vspace{4pt}

Case (2). If $\pi_j=b^j_{0,\tau}$, we have
\begin{align}
&r_j \cdot [\upsilon_j^{\tau}(j,W_j^{\mathbf{r}}(b^j_{0,\tau}), \rho)]^{-1}= 1+\frac{(1-p^j_{11})(\tau+1-\rho)}{\rho (b_{0,\tau}^j-b_{s}^j)+b_{s}^j} \nonumber \\
=&1+\frac{1-p^j_{11}}{b_{s}^j-b_{0,\tau}^j} \Big[1+\frac{b_{s}^j-(\tau+1)(b_{s}^j-b_{0,\tau}^j)}{\rho(b_{s}^j-b_{0,\tau}^j)-b_{s}^j}\Big]\label{eq:r_i_2}.
\end{align}

When $\tau>\tau_0$, it can be derived that the numerator $b_{s}^j-(\tau+1)(b_{s}^j-b_{0,\tau}^j)$ inside (\ref{eq:r_i_2}) is positive. Therefore, $\upsilon_j^{\tau}(j,W_j^{\mathbf{r}}(b^j_{0,\tau}), \rho)$ strictly increases with $\rho$ in this case.
\vspace{3pt}

Case (3). If $\pi_j=b^j_s$,
\begin{align}
r_j\cdot [\upsilon_j^{\tau}(j,W_j^{\mathbf{r}}(b^j_{s}), \rho)]^{-1}&=\frac{1}{b^j_s }\Big(\tau (1-p^j_{11})+\frac{1-p^j_{11}}{\rho}+b^j_s\Big)\nonumber
\end{align}

It is then clear from the above expression that $\upsilon^{\tau}_j(j,W_j^{\mathbf{r}}(b^j_{s}),\rho)$ strictly increases with $\rho$ in this case.

Now consider fixed $\rho$. For $\upsilon_j^{\tau}(W_j^{\mathbf{r}}(b^j_{0,h}), \rho)$ and $\upsilon_j^{\tau}(j,W_j^{\mathbf{r}}(b^j_{0,h+1}), \rho)$ with $h+1\leq\tau$, we have
\begin{align}\nonumber
\upsilon_j^{\tau}(j,W_j^{\mathbf{r}}(b^j_{0,h}), \rho)&\geq \upsilon_j^{\tau}(j,W_j^{\mathbf{r}}(b^j_{0,h}),0){=}\upsilon_j^{\tau}(j,W_j^{\mathbf{r}}(b^j_{0,h+1}),1)\nonumber\\
&\geq\upsilon_j^{\tau}(j,W_j^{\mathbf{r}}(b^j_{0,h+1}),\rho),\nonumber
\end{align}
where the first and last inequality is from case (1) we have just proven. The first equality is from expression~(\ref{eq:act_rew}). Since $\upsilon_j^{\tau}(j,W_j^{\mathbf{r}}(b^j_{0,h}), \rho)= \upsilon_j^{\tau}(j,W_j^{\mathbf{r}}(b^j_{0,h}),0)$ only if $\rho=0$, and $\upsilon_j^{\tau}(j,W_j^{\mathbf{r}}(b^j_{0,h+1}), 1)=\upsilon_j^{\tau}(j,W_j^{\mathbf{r}}(b^j_{0,h+1}),\rho)$ only if $\rho=1$. We hence have $\upsilon_j^{\tau}(j,W_j^{\mathbf{r}}(b^j_{0,h}), \rho)>\upsilon_j^{\tau}(j,W_j^{\mathbf{r}}(b^j_{0,h+1}),\rho)$ strictly. Following a similar derivation, we have $\upsilon_j^{\tau}(j,W_j^{\mathbf{r}}(b^j_{0,\tau}), \rho)>\upsilon_j^{\tau}(j,W_j^{\mathbf{r}}(b^j_{s}), \rho)$. Therefore the monotonicity property in part (i) holds for user $j$ with randomized transmission. The monotonicity result easily extends to user $i\neq j$ where there is no longer randomization in scheduling user $i$.

\vspace{4pt}

\noindent(ii) We proceed to prove part (ii) by first establishing the statement when $j_1=j_2=j$, $\omega_1=\omega_2=\omega$.\vspace{5pt}

Case (1). If $\omega=W_j^{\mathbf{r}}(b^j_{0,h})$ and $h<\tau$, from Lemma~\ref{lemma:alpha_tau}(i) and (\ref{eq:act_rew}) we have that
\begin{align}
&\upsilon_j^{\tau}(j,\omega,\rho)=r_j\Big[\alpha_j^{\tau}(j,b_{0,h}^j, \rho)+\nonumber\\
&\hspace{0.2in}+\frac{-(1-p^j_{11})}{\rho b_{0,h}^j+(1-\rho)b_{0,h+1}^j+(1-p^j_{11})(h+1-\rho)} \Big].\label{eq:R_A}
\end{align}

Case (2). If $\omega=W_j^{\mathbf{r}}(b^j_{0,\tau})$, we have
\begin{align}
&\upsilon_j^{\tau}(j,\omega,\rho)=r_j\Big[\frac{\rho b_{0,\tau}^j+(1-\rho)b_{s}^j}{\rho b_{0,\tau}^j+(1-\rho)b_{s}^j+(1-p^j_{11})(\tau+1-\rho)} \Big]\nonumber\\
=&r_j\Big[\alpha_j^{\tau}(j,\omega,\rho){+}\frac{-(1-p^j_{11})}{\rho b_{0,\tau}^j\hspace{-2pt}{+}(1{-}\rho)b_{s}^j{+}(1{-}p^j_{11})(\tau{+}1{-}\rho)} \Big].\label{eq:R_A_tau}
\end{align}

Case (3) If $\omega=W_j^{\mathbf{r}}(b^j_{s})$, we have
\begin{align}
&\upsilon_j^{\tau}(j,\omega,\rho)=r_j \Big[\frac{b^j_s \rho}{\tau \rho (1-p^j_{11})+(1-p^j_{11})+\rho b^j_s}\Big] \nonumber\\
=&r_j \Big[\alpha_j^{\tau}(j,\omega,\rho)+\frac{-\rho (1-p^j_{11})}{\tau \rho (1-p^j_{11})+(1-p^j_{11})+\rho b^j_s}\Big].\label{eq:R_A_bs}
\end{align}

Case (4). If $\omega>W_j^{\mathbf{r}}(b^j_s)$, since $\upsilon_j^{\tau}(j,\omega, \rho)=\alpha_j^{\tau}(j,\omega,\rho)=0$, the statement holds trivially.
\vspace{5pt}

Note that, in the above Case (1)-(3), the second summand in (\ref{eq:R_A})-(\ref{eq:R_A_bs}) decreases with the randomization parameter $\rho$. Since, from Lemma~\ref{lemma:alpha_tau}(ii) and part (i), both $\alpha_j^{\tau}(j,\omega, \rho)$ and $\upsilon_j^{\tau}(j,\omega,\rho)$ increase with $\rho$, we have for any $\rho_1>\rho_2$,
\begin{align}
0{\leq}\upsilon_j^{\tau}(j,\omega,\rho_1){-}\upsilon_j^{\tau}(j,\omega,\rho_2){\leq} r_j\big[\ \alpha_j^{\tau}(j,\omega,\rho_1){-}\alpha_j^{\tau}(j,\omega,\rho_2)\big]. \nonumber
\end{align}

We also have $\upsilon_i^{\tau}(j,\omega,\rho_1)=\upsilon_i^{\tau}(j,\omega,\rho_1)$ and $\alpha_i^{\tau}(j,\omega,\rho_1)=\alpha_i^{\tau}(j,\omega,\rho_1)$ for $i\neq j$ since there is no randomization associated with user $i$. Therefore, for all user $i$,
\begin{align}
0{\leq}\upsilon_i^{\tau}(j,\omega,\rho_1){-}\upsilon_i^{\tau}(j,\omega,\rho_2){\leq} r_i\big[\ \alpha_i^{\tau}(j,\omega,\rho_1){-}\alpha_i^{\tau}(j,\omega,\rho_2)\big]. \label{eq:vij}
\end{align}

Next consider when $i<j$,
\begin{align}
0=&\upsilon_i^{\tau}(j,\omega,\rho){-}\upsilon_i^{\tau}(i,\omega,0)\nonumber\\
=&r_i\cdot\big[\ \alpha_i^{\tau}(i,\omega,1){-}\alpha_i^{\tau}(i,\omega,0)\big]\nonumber\\
=&r_i\cdot\big[\ \alpha_i^{\tau}(i,\omega,\rho){-}\alpha_i^{\tau}(i,\omega,0)\big].\label{eq:ji1}
\end{align}

When $i=j$, from (\ref{eq:vij}) we have
\begin{align}
\upsilon_i^{\tau}(j,\omega,\rho){-}\upsilon_i^{\tau}(i,\omega,0){\leq} r_i\big[\ \alpha_i^{\tau}(j,\omega,\rho){-}\alpha_i^{\tau}(i,\omega,0)\big].\label{eq:ji2}
\end{align}

When $i>j$,from (\ref{eq:vij}) we have
\begin{align}
&\upsilon_i^{\tau}(j,\omega,\rho){-}\upsilon_i^{\tau}(i,\omega,0)\nonumber\\
=&\upsilon_i^{\tau}(i,\omega,1){-}\upsilon_i^{\tau}(i,\omega,0)\nonumber\\
\leq&r_i\big[\ \alpha_i^{\tau}(i,\omega,1){-}\alpha_i^{\tau}(i,\omega,0)\big]\nonumber\\
\leq&r_i\big[\ \alpha_i^{\tau}(i,\omega,\rho){-}\alpha_i^{\tau}(i,\omega,0)\big].\label{eq:ji3}
\end{align}

Therefore, from (\ref{eq:ji1})-(\ref{eq:ji3}), we have
\begin{align}
\upsilon_i^{\tau}(j,\omega,\rho_1){-}\upsilon_i^{\tau}(i,\omega,0)\leq r_i\big[\ \alpha_i^{\tau}(i,\omega,1){-}\alpha_i^{\tau}(i,\omega,0)\big]\label{eq:ji_final}
\end{align}

Similarly, we have
\begin{align}
\upsilon_i^{\tau}(i,\omega,1){-}\upsilon_i^{\tau}(j,\omega,\rho)\leq r_i\big[\ \alpha_i^{\tau}(i,\omega,1){-}\alpha_i^{\tau}(j,\omega,\rho)\big].\label{eq:ij_final}
\end{align}

Now consider the case when $\omega_1\neq\omega_2$. Suppose $\omega_1=W_i^{\mathbf{r}}(b^i_{0,h_1})$ and $\omega_2=W_i^{\mathbf{r}}(b^i_{0,h_2})$ with $h_1<h_2\leq\tau$.
\begin{align}
&\Big|\upsilon_i^{\tau}(j_1,W_i^{\mathbf{r}}(b^i_{0,h_1}), \rho_1){-}\upsilon_i^{\tau}(j_2,W_i^{\mathbf{r}}(b^i_{0,h_2}),\rho_2)\Big|\nonumber\\
\leq& \Big|\upsilon_i^{\tau}(j_1,W_i^{\mathbf{r}}(b^i_{0,h_1}), \rho_1){-}\upsilon_i^{\tau}(i,W_i^{\mathbf{r}}(b^i_{0,h_1}), 0)\nonumber\\
&\hspace{0.03in}+\sum_{h_1<h<h_2}\Big[\upsilon_i^{\tau}(i,W_i^{\mathbf{r}}(b^i_{0,h}), 1)-\upsilon_i^{\tau}(i,W_i^{\mathbf{r}}(b^i_{0,h}), 0)\Big]\nonumber\\
&+\upsilon_i^{\tau}(i,W_i^{\mathbf{r}}(b^i_{0,h_2}),1)-\upsilon_i^{\tau}(j_2,W_i^{\mathbf{r}}(b^i_{0,h_2}),\rho_2)\Big| \nonumber\\
\leq& r_i\Big|\alpha_i^{\tau}(j_1,W_i^{\mathbf{r}}(b^i_{0,h_1}), \rho_1){-}\alpha_i^{\tau}(i,W_i^{\mathbf{r}}(b^i_{0,h_1}), 0)\nonumber\\
&\hspace{0.03in}+\sum_{h_1<h<h_2}\Big[\alpha_i^{\tau}(i,W_i^{\mathbf{r}}(b^i_{0,h}), 1)-\alpha_i^{\tau}(i,W_i^{\mathbf{r}}(b^i_{0,h}), 0)\Big]\nonumber\\
&\hspace{0.1in}+\alpha_i^{\tau}(i,W_i^{\mathbf{r}}(b^i_{0,h_2}),1)-\alpha_i^{\tau}(j_2W_i^{\mathbf{r}}(b^i_{0,h_2}),\rho_2)\Big| \nonumber\\
=& r_i\Big|\alpha_i^{\tau}(j_1,W_i^{\mathbf{r}}(b^i_{0,h_1}), \rho_1){-}\alpha_i^{\tau}(j_2,W_i^{\mathbf{r}}(b^i_{0,h_2}),\rho_2)\Big|\nonumber\\
=& r_i\Big|\alpha_i^{\tau}(j_1,\omega_1, \rho_1){-}\alpha_i^{\tau}(j_2,\omega_2,\rho_2)\Big|,\nonumber
\end{align}
where the first inequality is because $\upsilon_i^{\tau}(i,W_i^{\mathbf{r}}(b^i_{0,h}),0)=\upsilon_i^{\tau}(i,W_i^{\mathbf{r}}(b^i_{0,h+1}),1)$ and $\alpha_i^{\tau}(i,W_i^{\mathbf{r}}(b^i_{0,h}),0)=\alpha_i^{\tau}(i,W_i^{\mathbf{r}}(b^i_{0,h+1}),1)$, which can be observed from (\ref{eq:act_rew}) and Lemma~\ref{lemma:alpha_tau}(i). The second equality is from (\ref{eq:ji_final})-(\ref{eq:ij_final}). For other combinations of $\omega_1$ and $\omega_2$, the proof holds similarly. Part (ii) thus holds.

\bibliographystyle{IEEEbib}
\bibliography{Bib}

\end{document}